\documentclass[prb,showpacs]{revtex4}
\usepackage{graphics}
\usepackage{graphicx}% Include figure files
\usepackage{dcolumn} % Align table columns on decimal point
\usepackage{bm}
\usepackage{float}
\usepackage{epsfig}
\begin{document}
\def\rhov{{\mbox{\boldmath{$\rho$}}}}
\def\tauv{{\mbox{\boldmath{$\tau$}}}}
\def\Deltav{{\mbox{\boldmath{$\Delta$}}}}
\def\Lambdav{{\mbox{\boldmath{$\Lambda$}}}}
\def\Thetav{{\mbox{\boldmath{$\Theta$}}}}
\def\Psiv{{\mbox{\boldmath{$\Psi$}}}}
\def\Phiv{{\mbox{\boldmath{$\Phi$}}}}
\def\sigmav{{\mbox{\boldmath{$\sigma$}}}}
\def\xiv{{\mbox{\boldmath{$\xi$}}}}
\def\oh{{\scriptsize 1 \over \scriptsize 2}}
\def\ot{{\scriptsize 1 \over \scriptsize 3}}
\def\of{{\scriptsize 1 \over \scriptsize 4}}
\def\tf{{\scriptsize 3 \over \scriptsize 4}}
\title{Symmetry Analysis for the Ruddlesden-Popper Systems, 
Ca$_3$Mn$_2$O$_7$ and Ca$_3$Ti$_2$O$_7$}

\author{A. B. Harris}

\affiliation{Department of Physics and Astronomy, University of Pennsylvania,
Philadelphia, OA 19104}

\date{\today}

\begin{abstract}
We perform a symmetry analysis of the zero-temperature instabilities of the
tetragonal phase of Ca$_3$Mn$_2$O$_7$ and Ca$_3$Ti$_2$O$_7$ which is stable
at high temperature.  We introduce order parameters to characterize each of
the possible lattice distortions in order to construct a Landau free energy
which elucidates the proposed
group-subgroup relations for structural transitions in these systems.
We include the coupling between the unstable distortion modes and the 
macroscopic strain tensor.  We also analyze the symmetry of the
dominantly antiferromagnetic ordering which allows weak ferromagnetism.
We show that in this phase the weak ferromagnetic moment and the spontaneous
ferroelectric polarization are coupled, so that rotating one of these
ordering by applying an external electric or magnetic field one can rotate
the other ordering.  We discuss the number of different domains (including
phase domains) which exist in each of the phases and indicate how
these may be observed.
\end{abstract}
\maketitle

\section{INTRODUCTION}

\begin{figure}[h!]
\begin{center}
\includegraphics[width=8.6 cm]{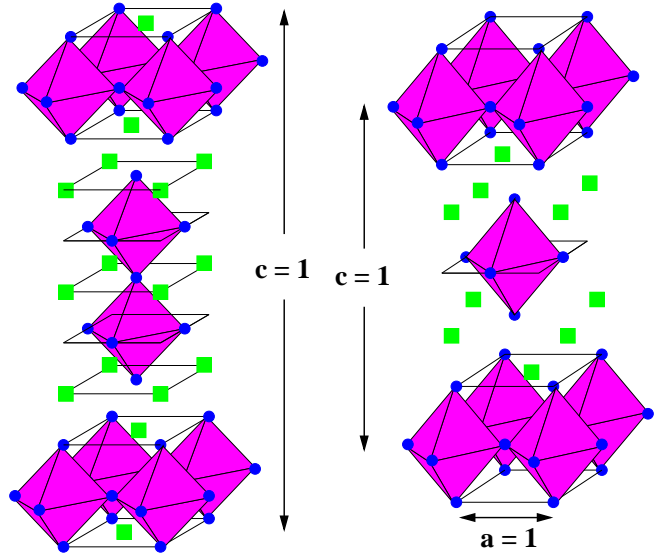}
\caption{\label{FIG1} (Color online) The unit cell of the RP ($n=2$) system
(left) and the ($n=1$) system (right).}
\end{center}
\end{figure}

Ruddlesden-Popper (RP) systems [\onlinecite{RP}] are compounds of the form
A$_{2+n}$B$_{1+n}$C$_{4+3n}$, where $n$ is an integer, and the valences of
the ions are usually A $=+2$, B  $=+4$ and C (oxygen) $=-2$.
Systems like SrTiO$_3$ can be regarded as being $n=\infty$.  
At high temperatures $T \approx 800$K, their crystal structure is
tetragonal, consisting of $n$-layer units, each layer consisting of 
vertex sharing oxygen octahedra at whose center sit a B ion,
as shown in Fig. \ref{FIG1}.
(For an alternative illustration, see Ref. \onlinecite{ARIAS}.
As the temperature is lowered these systems can undergo structural
phase transitions into orthorhombic
structures.[\onlinecite{STRUC1,STRUC2,STRUC3,STRUC4,STRUC5}] 
These structural transitions usually involve reorientation of oxygen
octahedra, a subject which has a long history,[\onlinecite{ARIAS}]
of which the most relevant references for the present paper are
those[\onlinecite{STR1,STR2,STR3}] which deal with the general
symmetry aspects of these transitions. One reason
for the continuing interest in octahedral reorientations is because
they are important for many interesting electronic properties, such as
high-$T_c$ superconductivity,[\onlinecite{BM}] 
colossal magnetoresistance,[\onlinecite{CMR}],
metal-insulator transitions,[\onlinecite{ARG}]
and magnetic ordering.[\onlinecite{LOB}]

Here we will focus on the $n=2$ RP systems, Ca$_3$Mn$_2$O$_7$ (CMO) and
Ca$_3$Ti$_2$O$_7$ (CTO), which present less complex scenarios than the
$n=1$ systems. It seems probable[\onlinecite{STRUC5,LAB}]
that well above room temperature
the crystal structure of CMO is that of the tetragonal space group
I4/mmm, \# 139 (space group numbering is
that of Ref. \onlinecite{ITC}), and at room temperature that of
space group Cmc2$_1$ (\# 36).[\onlinecite{GUIB}] The
theory of isotropy subgroups[\onlinecite{UTAH}] strongly forbids a
direct transition from I4/mmm to Cmc2$_1$. In fact, first
principles calculations on these systems by Benedek and Fennie
(BF) [\onlinecite{NATURE}] and on other systems[\onlinecite{PEREZ}]
indicate that the transition from I4/mmm to Cmc2$_1$ should proceed
via an intermediate phase which is probably Cmcm (\#63),
consistent with Ref. \onlinecite{UTAH}.
Up to now no such intermediate phase has been observed for CMO or CTO.
Unlike the other phases, the Cmc2$_1$ phase does not possess
a center of inversion symmetry and is allowed to have a spontaneous
polarization. Recent measurements on ceramic Ca$_3$Mn$_2$O$_7$ find a clear
pyroelectric signal consistent with the onset of ferroelectric order
close to $T^*=280$K.[\onlinecite{LAWES}] Therefore $T^*$ is identified
as the temperature at which Cmc2$_1$ appears.  Since this ferroelectric
transition seems
to be a continuous and well-developed one and since a direct continuous
transition between I4/mmm and Cmc2$_1$ is inconsistent with Landau
theory,[\onlinecite{UTAH}] the seemingly inescapable conclusion is that
the phase for $T$ slightly greater than $T^*$ is {\it not} I4/mmm, but
is some other phase which does not allow a spontaneous polarization. 
Thus the phase at temperature just above $T=280$K may be the long
sought for Cmcm phase.  In fact, the Cmcm phase has been observed in the
isostructural compounds LaCa$_2$Mn$_2$O$_7$[\onlinecite{GREEN}]
and Bi$_{0.44}$Ca$_{2.56}$Mn$_2$O$_7$[\onlinecite{QIN}] at room temperature.
In view of the results of Refs. \onlinecite{STRUC5} and \onlinecite{LAB},
it is possible that the Cmcm phase may exist only over a narrow range
of temperature.  As the temperature is further lowered, an antiferromagnetic
phase is observed.[\onlinecite{JUNG,LOB}] In this phase, which appears at 
$T=115$K,[\onlinecite{LOB}] the antiferromagnetic order is accompanied
by weak ferromagnetism.[\onlinecite{JUNG,LOB}]

Theoretically, there have been efforts to understand systems 
like these from first principles calculations. For instance, Ref.
\onlinecite{WHANG} found the nearest neighbor exchange $J_{\rm nn}$
within a bilayer to be $J_{\rm nn}/k_{\rm B}=-39$K, giving a Curie-Weiss
$\Theta=-244$K, whereas Ref. \onlinecite{CARDOSO} found
$J_{\rm nn}/k_{\rm B}=200$K.  The former calculation agrees much
better with experiment[\onlinecite{FAW}] which gave $\theta=-465$K.
Both groups studied the electronic band structure but it was
not entirely clear what space group their calculations predicted.
More detailed information on the symmetry of the structures
comes from the first principles calculations of BF some of which
included spin-orbit interactions.  These calculations give a 
weak ferromagnetism of $0.18 \mu_{\rm B}$ per unit cell which is somewhat
smaller than $0.3 \mu_{\rm B}$ per {\it spin}.[\onlinecite{FOOT}]
However, inclusion of spin-orbit interactions enabled BF to
obtain the correct symmetry of the magnetoelectric behavior.
Here we discuss in detail the symmetry properties
of the various phases and experimental consequences such as the 
interactions between various order parameters (OP's) and the number 
and symmetry of the various domains which may be observed. Several of
these issues were discussed by BF, but a more complete and 
systematic analysis is given here.

Our approach to symmetry is similar to that of Ref. \onlinecite{PEREZ} in
connection with the Aurivillius compound SrBi$_2$Ta$_2$O$_9$ (SBTO):
we adopt the high-temperature tetragonal structure as the ``reference"
structure and analyze the instabilities at zero temperature which lead to
the lower temperature phases.  Although CMO and CTO do differ from SBTO,
their crystal symmetry is the same as SBTO and hence many of the results
we find here are similar to those for SBTO.  Here we emphasize some
of the experimental consequences of the symmetries we find (such as
the enumeration of the different possible structurally ordered domains)
and also we explore the nature of the macroscopic strains, the
ferroelectric polarization, the magnetic ordering, and the
coupling between structural distortions and these degrees of freedom.

Briefly, this paper is organized as follows.  In Sec. II we outline
the basic approach used to analyze the symmetry of the systems in question.
In Sec. III we give a symmetry analysis of the resulting phases which result
from the structural instabilities found by BF.  This analysis closely
parallels that of Perez-Mato {et al.}[\onlinecite{PEREZ}].  In Sec. IV
we discuss the second structural transition in which the other two
irreps condense to reach the Cmc2$_1$ phase.  In Sec. V
we discuss the symmetry of the magnetic ordering.  We also explore the 
coupling betweeen the distortions and the strains, the polarization, and
the magnetic ordering.
Throughout the paper we point out experiments which are needed in order
to remove crucial  gaps in our understanding of the structural phase
diagram of these systems. In Sec. VI we enumerate the possible domains
that can occur and briefly discuss their dynamics.  Our conclusions are
summarized in Sec. VII.

\section{SYMMETRY ANALYSIS}

Our symmetry analysis will be performed relative to the tetragonal I4/mmm
structure which is stable at high temperatures.  In this high-temperature 
reference structure one has atoms at their equilibrium positions
${\bf R}^{(0)} ( {\bf n}, \tau)$, where
\begin{eqnarray}
R_\alpha^{(0)} (n_1, n_2, n_3; \tau) &\equiv&
n_1 R_{1,\alpha} + n_2 R_{2,\alpha} + n_3 R_{3,\alpha} +
\tau_\alpha  \ ,
\nonumber \end{eqnarray}
where the lattice vectors are
\begin{eqnarray}
{\bf R}_1 &\equiv& (-1/2,1/2,1/2) \ , \ \ \ \
{\bf R}_2 \equiv (1/2,-1/2,1/2) \ , \nonumber \\
{\bf R}_3 &\equiv& (1/2,1/2,-1/2) \ , 
\nonumber \end{eqnarray}
where $\alpha$ labels Cartesian components,  real space coordinates
are expressed as fractions of lattice constants (so that, for instance,
${\bf R}_1$ denotes $(-a/2,a/2,c/2)$), and the number $\tau$ labels sites
within the unit cell at the position vector $\tauv$, as listed in Table 
\ref{basis}.  The associated tetragonal reciprocal lattice vectors are
\begin{eqnarray}
{\bf G_1} &=& (0,1,1) \ , \ \ \
{\bf G_2} = (1,0,1) \ , \ \ \
{\bf G_3} = (1,1,0) \ , \nonumber 
\end{eqnarray}
in reciprocal lattice units so that, for instance,
${\bf G_1}$ denotes $2 \pi (0,1/a,1/c)$.

\begin{figure}[h!]
\begin{center}
\includegraphics[width=8.6 cm]{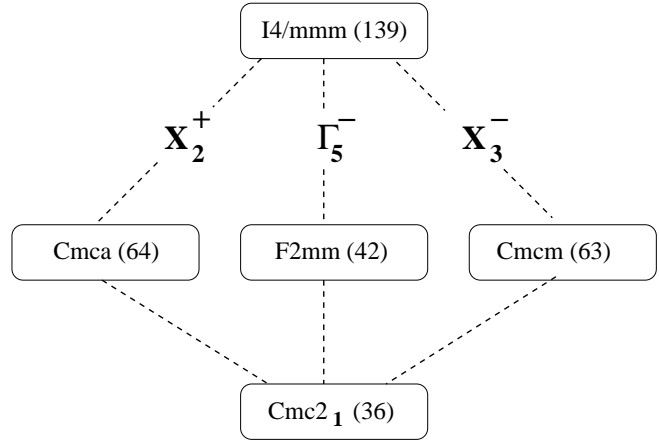}
\caption{\label{FIG2} (Color online) The group-subgroup structure 
arising from the first principles calculations.  The transitions from
I4/mmm are labeled with irrep that is condensing.  In the final
transition to Cmc2$_1$ we show below that the two remaining irreps
condense to that all three irreps have condensed to form Cmc2$_1$.
This diagram does not deal with magnetic ordering.}
\end{center}
\end{figure}

This work was stimulated by the first principles calculations of BF
on the instabilities at zero temperature of the reference tetragonal
system.  For CMO the instabilities at zero temperature occur
for the irreps $X_3^-$ and $X_2^+$ at the zone boundary $X$ points
which are ${\bf K}=(1/2,1/2,0)\equiv {\bf q}_1$ and 
${\bf K}=(1/2,-1/2,0)\equiv {\bf q}_2$.  (The superscript indicates
the parity under inversion about the origin.) Since $-{\bf q}_n$ is
equal (modulo a reciprocal lattice vector) to ${\bf q}_n$, the
vectors ${\bf q}_1$ and ${\bf q}_2$ exhaust the star of $X$.  Also
a zone center phonon of the two dimensional irrep $\Gamma_5^-$ is
nearly unstable. These results suggest the group-subgroup structure 
shown in Fig. \ref{FIG2}, which is similar to that for
SBTO.[\onlinecite{PEREZ}]
We therefore consider structures having distorted positions given by
\begin{eqnarray}
R_\alpha({\bf n}, \tau) &=& R_\alpha^{(0)} ({\bf n}, \tau)
+ u_\alpha (\tau) e^{i {\bf q} \cdot ( \sum_k {\bf R}_k n_k)}
\nonumber \\ && \
+ \sum_\beta e_{\alpha \beta} R_\beta^{(0)}({\bf n}, \tau) \ ,
\nonumber \end{eqnarray}
where ${\bf q}$ is the wave vector of the distortion mode,
${\bf u}(\tau)$ is the distortion taken from Table \ref{basis}, 
and $\epsilon_{\alpha \beta}$ is the macroscopic strain tensor.
The leading terms in the Landau expansion of the free energy of the 
distorted structure relative to the reference tetragonal structure 
will have terms quadratic in the strains $e_{\alpha \beta}$
and the microscopic displacements ${\bf u}(\tau )$ within the unit cell.
The fact that the free energy has to be invariant under all the
symmetry operations of the ``vacuum" (i. e. the reference tetragonal
structure) restricts the microscopic displacements to be a linear
combination of the basis vectors of the irrep in question.  The
basis functions are listed in Table \ref{basis} and the
representation matrices for the generators of the irreps are given in Table
\ref{IRREP}[\onlinecite{FN1}] in terms of the Pauli matrices
\begin{eqnarray}
\sigmav_x &=& \left[ \begin{array} {cc} 0&1\\1&0\\ \end{array} \right]
\hspace{0.2 in}
\sigmav_y = \left[ \begin{array} {cc} 0&-i\\i&0\\ \end{array} \right]
\hspace{0.2 in}
\sigmav_z = \left[ \begin{array} {cc} 1&0\\0&-1\\ \end{array} \right] \ .
\nonumber \end{eqnarray}

\begin{table} [h!]
\caption{\label{basis} Basis functions $\Psi_1^{(Z)}({\bf R},\tau)
= \Psi^{(Z)}_{1,n}(\tau) \cos ({\bf q}\cdot {\bf R})$
and $\Psi_2^{(Z)}({\bf R},\tau)=\Psi^{(Z)}_{2,n}(\tau)
\cos({\bf q} \cdot {\bf R})$ for the distortion vector under irrep $Z$,
where $Z=5,3,2$ indicates irrep $\Gamma_5^-$, $X_3^-$, and $X_2^+$,
respectively, and $n=x,y,z$ labels the components.  The $\Psi$'s
are normalized according to Eq. (\ref{NORM}).  For the $X$ irreps
the wave vector of $\Psi_{k,n}^{(X)}({\bf R}, \tau)$ is ${\bf q}_k$.
For $\Gamma_5^-$ the wave vector is ${\bf q}=0$. For each site $\tau$
we give the three components ($n=1,2,3)$ of the vector displacement.
The values of the displacements in this table are not restricted by symmetry.
The values of the structure parameters for CMO, taken from Ref.
\onlinecite{STRUC5}, are $\rho=0.311$, $\xi=0.100$, $\chi=0.205$,
and $\tau=0.087$.}
\vspace{0.2 in}
\begin{tabular} {|| c | c || c c c | c c c ||c c c | c c c ||c c c 
| c c c ||}
\hline \hline
$\tau$ & $\tauv$ & \multicolumn{3} {|c|} { $\Psi_{1,n}^{(5)}(\tau)$} &
\multicolumn{3}{|c|} {$\Psi_{2,n}^{(5)}(\tau)$} &
\multicolumn{3}{|c|} {$\Psi_{1,n}^{(3)}(\tau)$} &
\multicolumn{3}{|c|} {$\Psi_{2,n}^{(3)}(\tau)$} &
\multicolumn{3}{|c|} {$\Psi_{1,n}^{(2)}(\tau)$} &
\multicolumn{3}{|c|} {$\Psi_{2,n}^{(2)}(\tau)$} \\
\hline
\multicolumn{20} {|c||} {A sites} \\ \hline
1& $(0,0,\rho+1/2) $ & $u$  & 0 & 0 & 0 & $u$  & 0 
& $a$ & $-a$ & 0 & $-a$ & $-a$ & 0 & 0 & 0 & 0 & 0 & 0 & 0 \\
2& \ \ $(0,0,-\rho+1/2)$ \ \  &$u$&0&0 &0&$u$&0
& $a$ & $-a$ & 0 & $-a$ & $-a$ & 0 & 0 & 0 & 0 & 0 & 0 & 0 \\
3& $(0,0,1/2)   $ & $v$  & 0 & 0 & 0 & $v$  & 0
& $b$ & $-b$ & 0 & $-b$ & $-b$ & 0 & 0 & 0 & 0 & 0 & 0 & 0 \\
\hline
\multicolumn{20} {|c||} {B sites} \\ \hline
4& $(0,0,\xi)   $ & $w$  & 0 & 0 & 0 & $w$  & 0
& $c$ & $-c$ & 0 & $-c$ & $-c$ & 0 & 0 & 0 & 0 & 0 & 0 & 0 \\
5& $(0,0,-\xi)  $ & $w$ & 0 & 0 & 0 &$w$  & 0
& $c$ & $-c$ & 0 & $-c$ & $-c$ & 0 & 0 & 0 & 0 & 0 & 0 & 0 \\
\hline
\multicolumn{20} {|c||} {O sites} \\ \hline
6 & $(0,0,0)     $ & $x$  & 0 & 0 & 0 & $x$  & 0
& $d$ & $-d$ & 0 & $-d$ & $-d$ & 0 & 0 & 0 & 0 & 0 & 0 & 0 \\
7 & $(0,0,\chi)  $ & $y$  & 0 & 0 & 0 & $y$  & 0
& $e$ & $-e$ & 0 & $-e$ & $-e$ & 0 & 0 & 0 & 0 & 0 & 0 & 0 \\
8 & $(0,0,-\chi) $ & $y$ & 0 & 0 & 0 &$y$  & 0
& $e$ & $-e$ & 0 & $-e$ & $-e$ & 0 & 0 & 0 & 0 & 0 & 0 & 0 \\
9 & $(0,1/2,\tau)$ & $z_1$ & 0 &0  &0  & $z_2$ &0
& $0$ & $0$ & $f$ & $0$ & $0$ & $f$ & $-h$ & $-g$ & 0 & $-h$ & $g$ & 0 \\
10 & $(0,1/2,-\tau)$ & $z_1$  &0 &0 & 0 & $z_2$ &0
& $0$ & $0$ & $-f$ & $0$ & $0$ & $-f$ & $-h$ & $-g$ & 0 & $-h$ & $g$ & 0 \\
11 & $(1/2,0,\tau)$ & $z_2$  &0 &0 &0 & $z_1$ &0
& $0$ & $0$ & $-f$ & $0$ & $0$ & $f$ & $g$ & $h$ & 0 & $-g$ & $h$ & 0 \\
12 & $(1/2,0,-\tau)$ & $z_2$  &0 &0 &0 & $z_1$ &0
& $0$ & $0$ & $f$ & $0$ & $0$ & $-f$ & $g$ & $h$ & 0 & $-g$ & $h$ & 0 \\
\hline \hline
\end{tabular}
\end{table}
\begin{table} [h!]
\caption{\label{IRREP} Representation matrices $M^{(3)}$, $M^{(2)}$, and
$M^{(5)}$ for the generators of the irreps $X_3^-$, $X_2^+$, and $\Gamma_5^-$,
respectively.  Here ${\bf r}'={\cal O}{\bf r}$ and the $\sigmav$'s are
the Pauli matrices.  These matrices are related to those of 
Ref. \onlinecite{UTAH}[\onlinecite{UTAH2}] by a unitary transformation
which, for $\Gamma_5^-$ takes $(\sigmav_x, \sigmav_y, \sigmav_z)$
into $(-\sigmav_x, -\sigmav_y, \sigmav_z)$.  For $X_2^+$ and $X_3^-$
the unitary transformation takes ($\sigmav_x, \sigmav_y, \sigmav_z)$
into $(-\sigmav_z, \sigmav_y, -\sigmav_x)$. Unlike Ref. \onlinecite{UTAH}
we choose a representation in which the matrices representing the
translations ${\bf T}_n$ are diagonal.}
\vspace{0.2 in}
\begin{tabular} {|| c | c c c | c c c ||}
\hline \hline
${\cal O}=$ & ${\cal R}_4$ & $m_d$ & $m_z$ & ${\bf T}_1$ & ${\bf T}_2$ &
${\bf T}_3$ \\
${\bf r}' = $ &\ \  $(\overline y, x,z)$\ \ &\ \ $(y,x,z)$\ \ &
\ \ $(x,y, \overline z)$\ \ & \ \ $(x+1,y,z) $\ \ &\ \ $(x,y+1,z)$\ \ &
\ \ $(x+\frac{1}{2}, y+ \frac{1}{2}, z-\frac{1}{2})$\ \ \\
\hline
${\bf M}^{(3)}({\cal O})=$ & $-i \sigmav_y$ & $-\sigmav_z$ & ${\bf 1}$ &
$- {\bf 1}$ & $ -{\bf 1}$ & $-\sigmav_z$ \\
\hline
${\bf M}^{(2)}({\cal O})=$ & $\sigmav_x$ & $- {\bf 1}$ & ${\bf 1}$ &
$- {\bf 1}$ & $- {\bf 1}$ & $-\sigmav_z$ \\
\hline
${\bf M}^{(5)}({\cal O})=$ & $i \sigmav_y$ & $\sigmav_x$ & ${\bf 1}$ &
${\bf 1}$ & ${\bf 1}$ & ${\bf 1}$ \\
\hline \hline
\end{tabular}
\end{table}

Using the basis functions  $\Psi_1^{(Z)}$ and $\Psi_2^{(Z)}$ listed
in Table \ref{basis} one can check that the matrices of Table \ref{IRREP}
do form a representation, so that for an operator ${\cal O}$ we have
\begin{eqnarray}
{\cal O} \left[ \begin{array} {c} \Psi_1^{(Z)} \\ \Psi_2^{(Z)} \\ \end{array} \right]
= \left[ \begin{array} {c c } M_{11}({\cal O}) &
M_{12}({\cal O}) \\ M_{21}({\cal O}) & M_{22}({\cal O})
\\ \end{array} \right] \left[ \begin{array} {c} \Psi_1^{(Z)} \\ \Psi_2^{(Z)} \\
\end{array} \right] \ ,
\nonumber \end{eqnarray}
where the $\Psi$'s are normalized:
\nonumber \begin{eqnarray}
\sum_{n \tau} \Psi^{(Z)}_{k,n} (\tau)^2 = 1 {\AA}^2  \ .
\label{NORM} \end{eqnarray}
We now introduce OP's $Q_1^{(Z)}$ and $Q_2^{(Z)}$ as amplitudes
of these normalized distortions, so that a distortion $\Phi_Z$ of the 
irrep Z can
be written as $\Phi_Z = Q_1^{(Z)} \Psi_1^{(Z)} + Q_2^{(Z)} \Psi_2^{(Z)}$ and
\begin{eqnarray}
{\cal O} \Phi_Z & \equiv & {\cal O} [ Q_1^{(Z)} \Psi_1^{(Z)}
+ Q_2^{(Z)} \Psi_2^{(Z)} ] \nonumber \\  &=&
[M_{11}^{(Z)}({\cal O}) Q_1^{(Z)} + M_{21}^{(Z)}({\cal O}) Q_2^{(Z)}]
\Psi_1^{(Z)}
\nonumber \\ && + [M_{12}^{(Z)}({\cal O}) Q_1^{(Z)}
+ M_{22}^{(Z)}({\cal O}) Q_2^{(Z)}] \Psi_2^{(Z)}  \ .
\nonumber \end{eqnarray}

We now interpret this as defining how the order parameters transform
when the $\Psi_n^{(Z)}$ are regarded as fixed. Thus we have
\begin{eqnarray}
{\cal O} \left[ \begin{array} {c} Q_1^{(Z)} \\ Q_2^{(Z)} \end{array} \right]
&=& \left[ \begin{array} {cc} M_{11}^{(Z)}({\cal O}) & M_{21}^{(Z)}({\cal O})
\\ M_{12}^{(Z)}({\cal O}) & M_{22}^{(Z)}({\cal O}) \\ \end{array}
\right] \left[ \begin{array} {c} Q_1^{(Z)} \\ Q_2^{(Z)} \\
\end{array} \right] \ .
\label{TRANSEQ} \end{eqnarray}
Note that the OP's transform according to the {\it transpose} of the
irrep matrices.

\section{Unstable Irreps}

\subsection{$X_3^-$}

\begin{figure}[h!]
\begin{center}
\includegraphics[width=16.0 cm]{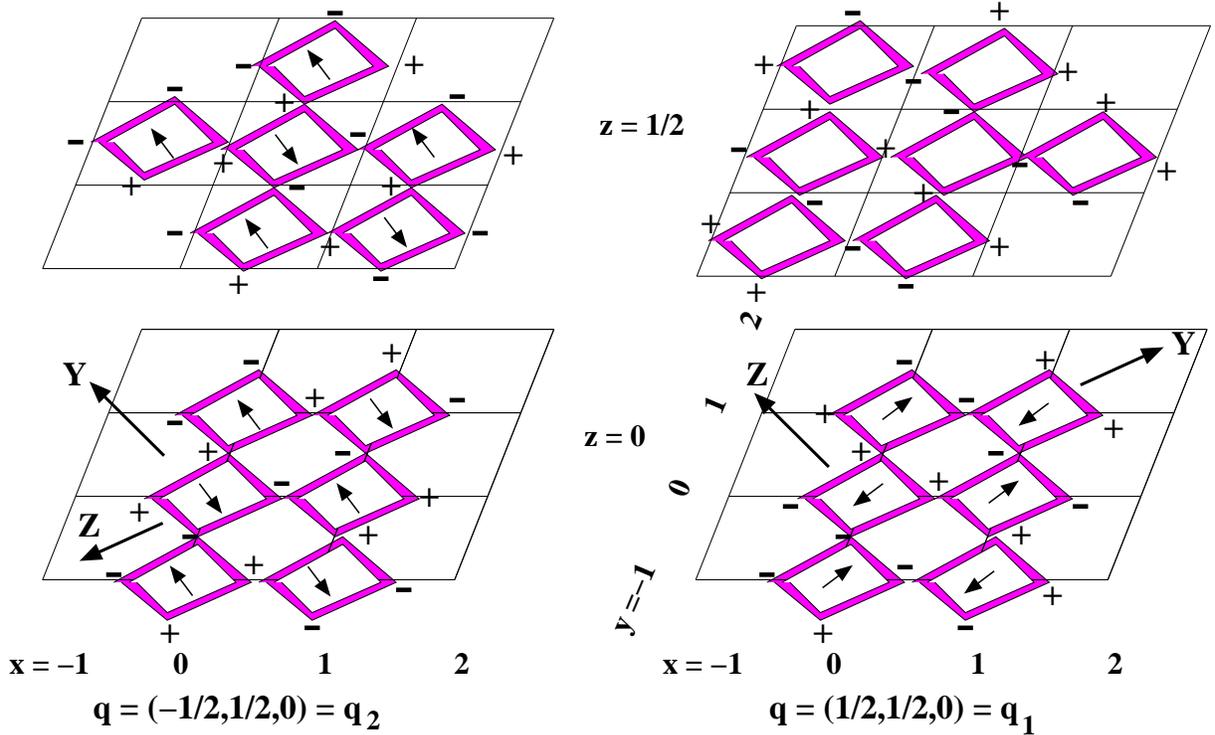}
\caption{\label{FIGX3} (Color online) Schematic diagram of the displacements 
$\Psi^{(3)}_1({\bf R},\tau)$, left and $\Psi^{(3)}_2({\bf R},\tau)$, right
of the oxygen octahedra.
All coordinates are in the parent tetragonal system.
The arrows represent displacements within the $x$-$y$ plane and the $+$ and $-$
signs, displacements collinear with the $z$-axis.  The arrows represent the
displacement of the apical oxygen at $(0,0,\chi)$,
which is the same as that at $(0,0,-\chi)$. The displacement of the shared
apical oxygen at $(0,0,0)$ is not shown.  The algebraic sign of the
displacements in the $z$ direction of the equatorial oxygen at $(x,y,\tau)$
in the unit cell is indicated.  The displacement of the oxygen at $(x,y,-\tau)$
is the negative of that at $(x,y,\tau)$.  The orthorhombic
axes for Cmcm are $X$, $Y$, and
$Z$.  The distortion of left panel is obtained from that of the right panel
by a 90$^{\rm o}$ rotation about the positive $z$-axis.}
\end{center}
\end{figure}

The irrep $X_3^-$ of the little group (of the wave vector) is one
dimensional.  However, since there are two wave vectors in the star
of $X$, we will follow Ref. \onlinecite{UTAH} and construct the two
dimensional irrep which incorporates both wave vectors in the star of 
${\bf X}$.  The resulting two dimensional matrices are given in
Table \ref{IRREP}.
From the basis functions for irrep $X_3^-$ given in Table \ref{basis},
one sees that the distortion can describe the alternating tilting
of the oxygen octahedra about a $(1,\overline 1 ,0)$ direction if
${\bf q}={\bf q}_1$ and about a $(1, 1 , 0)$ direction if
${\bf q}={\bf q}_2$, as shown in Fig. \ref{FIGX3}. We now
construct the form of the free energy when the distortion is given 
by[\onlinecite{FNA}]
\begin{eqnarray}
\Phi &=& Q_3^-({\bf q}_1) \Psi_1^{(3)} + Q_3^- ({\bf q}_2) \Psi_2^{(3)} \ .
\nonumber \end{eqnarray}

To construct the form of the free energy for this structure, note that
wave vector conservation requires that the free energy be a function of
$Q_3({\bf q}_1)^2$ and $Q_3({\bf q}_2)^2$ because $2 {\bf q}_1$ and
$2 {\bf q}_2$ are reciprocal lattice vectors but ${\bf q}_1 + {\bf q}_2$
is not a reciprocal lattice vector.  Then, using the irrep matrices given
in Table \ref{IRREP}, one can check that the free energy in terms
of the $X_3^-$ OP's must be of the form
\begin{eqnarray}
{\cal F }({\bf X}_3) &=& \frac{a}{2}(T-T_3) [Q_3({\bf q}_1)^2 
+ Q_3({\bf q}_2)^2] \nonumber \\ && \ + 
\frac{1}{4} u [Q_3({\bf q}_1)^2 + Q_3({\bf q}_2)^2 ]^2  \nonumber \\ && \
+ v [Q_3({\bf q}_1) Q_3({\bf q}_2)]^2  + {\cal O} (Q^6) \ ,
\label{F3-} \end{eqnarray}
where $a>0$ and $T_3$ is the temperature at which this irrep becomes active 
if it is the only relevant OP). The first principles calculations
of BF indicate that $u > 0$.  To treat the 
coupling between the $Q$'s and the strains we consider the strain-dependent
contribution to the free energy $F_{Q\epsilon}$. The term whereby the
$Q$'s induce a strain is linear in the strain and thus we write
$F_{Q\epsilon}$ as
\begin{eqnarray}
F_{Q\epsilon} &=& \frac{1}{2} \sum_{nm} c_{nm} \epsilon_n \epsilon_m
+ \sum_{klm} \gamma_{klm} \epsilon_{kl} Q_3^-({\bf q}_m)^2  
\nonumber \\ &\equiv&
\frac{1}{2} \sum_{nm} c_{nm} \epsilon_n \epsilon_m + V_{Q \epsilon} \ ,
\nonumber \end{eqnarray}
where the first term is in Voigt[\onlinecite{VOIGT}] notation, where
$1 \equiv x,x$, etc. and only $c_{11}$, $c_{12}$, $c_{13}$, $c_{33}$,
$c_{44}$, and $c_{66}$ are nonzero under tetragonal symmetry and
\begin{eqnarray}
V_{Q,\epsilon} &=& 
\alpha \epsilon_{xy}[Q_3^-({\bf q}_1)^2 - Q_3^-({\bf q}_2)^2 ]
+ [\beta (\epsilon_{xx} + \epsilon_{yy}) \nonumber \\ && \ + \gamma
\epsilon_{zz}] [Q_3^-({\bf q}_1)^2 + Q_3^-({\bf q}_2)^2 ] \ ,
\label{G3STRAIN} \end{eqnarray}
where $\alpha$, $\beta$, and $\gamma$ (and similarly below) are arbitrary
constants.  The sign of the shear deformation $\epsilon_{xy}$
depends on the sign of $Q_3^- ({\bf q}_1)^2  - Q_3^-({\bf q}_2)^2$, a result
similar to that given in Ref. \onlinecite{AXE}.

As the temperature is reduced through the value $T_3$, a distortion of
symmetry $X_3^-$ appears.  For $v>0$ either one, but not both,
of $Q_3({\bf q}_1)$ and
$Q_3({\bf q}_2)$ are nonzero, so that $\epsilon_{xy} \not= 0$ and a
detailed analysis shows that the resulting structure is Cmcm (\#63).
The first principles calculations of BF imply
that $v>0$, so, as indicated by Fig. \ref{FIG2},
this possibility is the one realized for the RP systems we have studied.
If $v$ had been negative, then we would have had a ``double-${\bf q}$"
state ({\it i. e.} a state simultaneously having two wave vectors) with
$|Q_3({\bf q}_1)|=|Q_3({\bf q}_2)|$, so
that $\epsilon_{xy}=0$ and the distorted structure would be the tetragonal
space group P4$_2$/mnm (\#136).  These two space
groups are listed in Ref. \onlinecite{UTAH} as possible subgroups of 
I4/mmm which can arise out of the irrep $X_3^-$. These
directions in OP-space [$Q_3({\bf q}_1)-Q_3({\bf q}_2)$-space] are
stable with respect to perturbations due to higher order terms
in Eq. (\ref{F3-}) which are anisotropic in OP-space as long as $T$
is close enough to $T_3$ so that these perturbations are sufficiently small.
We note that Ref. \onlinecite{UTAH} lists Pnnm (\#58) 
as an additional 
possible subgroup.  This subgroup would arise if $v$ were exactly  zero, 
in which case, even for $T$ arbitrarily close to $T_3$,
the OP's would be determined by higher order terms in Eq. (\ref{F3-})
and would then not be restricted to lie along a high symmetry direction
in OP space.  However, to realize this possibility
requires the accidental vanishing of the coefficient $v$ and the
analogous sixth order anisotropy.  This possibility should be
rejected unless additional control parameters, such as the pressure
or an electric field, are introduced which allow access to such a
multicritical point.[\onlinecite{STR3}]  As of this writing, the Cmcm
phase (or any other phase intermediate between I4/mmm and Cmc2$_1$) 
has not been observed, although it has been observed in the 
isostructural systems LaCa$_2$Mn$_2$O$_7$[\onlinecite{GREEN}] and 
Bi$_{0.44}$Ca$_{2.56}$Mn$_2$O$_7$[\onlinecite{QIN}]

\begin{figure}[h!]
\begin{center}
\caption{\label{ORTHO} (Color online) The tetragonal axes (lower case) and the
axes (capitals) we use to describe the orthorhombic phases.}
\vspace{0.2 in}
\includegraphics[width=4.8 cm]{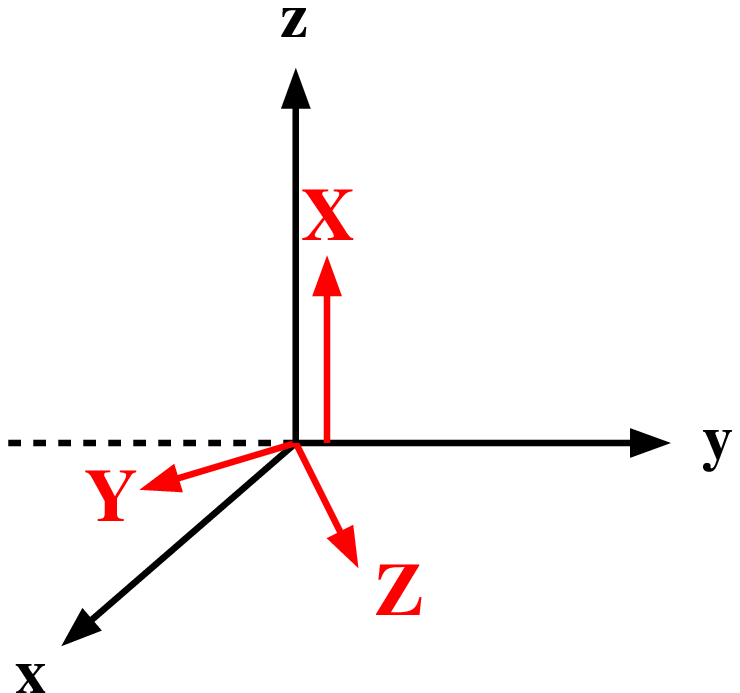}
\end{center}
\end{figure}

For future reference, it is
convenient to introduce orthorhombic axes, so we arbitrary
define them as in Fig. \ref{ORTHO}.  If we condense $Q_3({\bf q}_1)$,
the tilting is about $Y$, the tetragonal [1 $\overline 1$ 0]
direction, whereas if we condense $Q_3({\bf q}_2)$, the
tilting is about $Z$, the tetragonal [1 1 0] direction.
The sign of the OP indicates the sign of the tilting angle.
Note that there are four domains of ordering, depending
on which wave vector has condensed and the signs of the OP.
These correspond to the four equivalent initial orientations of
the tetragonal sample.  In Ni$_3$V$_2$O$_8$, such a domain
structure for two coexisting OP's was confirmed 
experimentally,[\onlinecite{DOMAIN}] and it would be nice to do
the same here.  In Sec. VI we give a more detailed
discussions of domains. Here we note that changing the sign
of the OP $Q_3({\bf q}_1)$ is equivalent to a unit translation
along $\hat x$ or $\hat y$.  Thus, as for an antiferromagnet,
the phase of the order parameter within a single domain
has no macroscopic consequences.

Since the Cmcm structure has a center of inversion symmetry, it can
not have a nonzero spontaneous polarization and an interaction 
$V_{Q,{\bf P}}$ which is linear in the polarization does not arise.
It may seem mysterious that when we introduce a distortion which is
odd under inversion, we still have a structure which is even under
inversion.  The point is that the distortion due to $X_3^-$ is
odd under inversion about the origin, but is even under inversion
{\it with respect to} $(1/2,0,0)$, about which point the parent
tetragonal structure also has a center of inversion symmetry.
To see the inversion symmetry of the distortion about $(1/2,0,0)$
from Fig. \ref{FIGX3}, note that inversion displaces the
arrows and reverses their direction.  Since a plus sign represents a
positive $z$ distortion at $z=\tau$ and a negative $z$ distortion at
$z=-\tau$, inversion takes a plus into a plus and a minus into a minus.

\subsection{$X_2^+$}

\begin{figure} [h!]
\begin{center}
\includegraphics[width=15.0 cm]{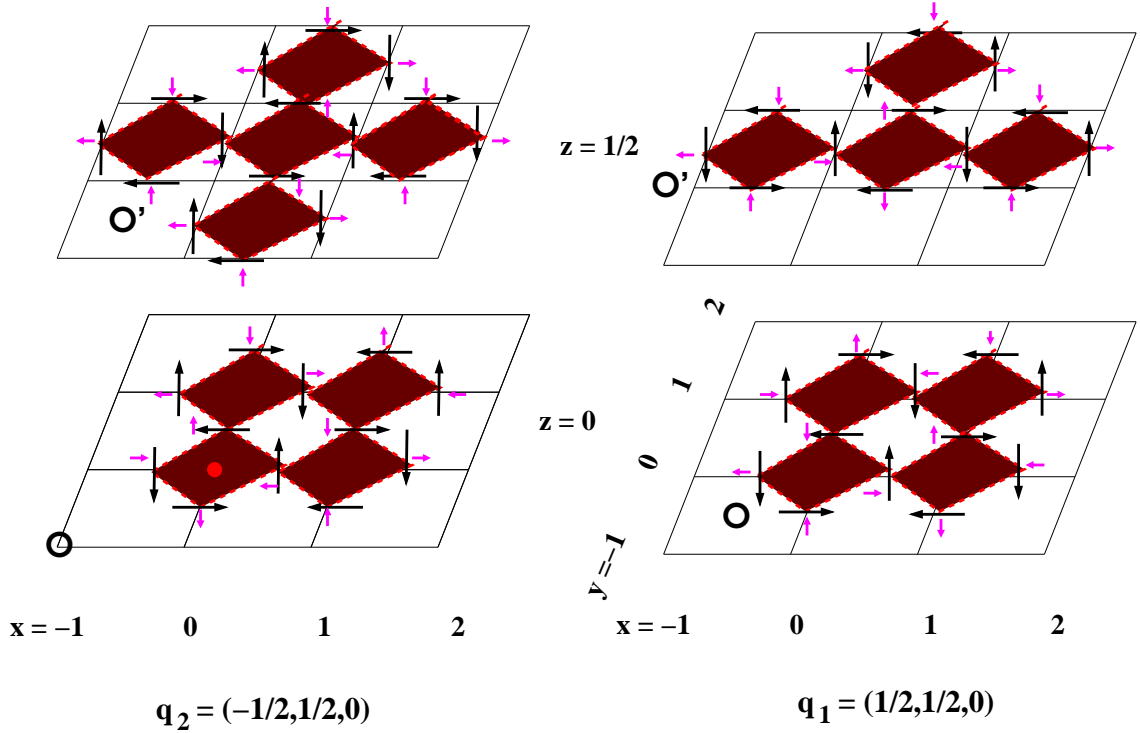}
\caption{\label{X2+} (Color online) Schematic diagram of the displacements
for $X_2^+$, which are confined to the $x$-$y$ plane.  All coordinates are
in the parent tetragonal system.  The large black arrows give the rotational
distortion. The much 
smaller anisotropic radial distortion (which is allowed in $X_2^+$)
is shown in magenta.  In both panels the distortions are shown for sites
in the planes $z=\pm \tau$ and $z = 1/2 \pm \tau$.
The distortion is an even function of $\tau$.
The distortion of the right panel is obtained from that of the left panel
by a 90$^{\rm o}$ rotation about the $z$-axis.  
The translation vectors in the $x$-$y$ plane are $(1,1,0)$ and $(1,-1,0)$.
In the left panel the translation vector $\Delta {\bf r}$ (which takes O 
into O' and obeys $\exp(i {\bf q} \cdot \Delta {\bf r} ) = 1$)
is $(1/2,1/2,1/2)$ and in the right panel it is $(-1/2,1/2,1/2)$.  These
two wave vectors thus give rise to the two settings of the side-centered 
orthorhombic lattice.}
\end{center}
\end{figure}

As in the case of $X_3^-$ we construct the two dimensional irrep which
incorporates both wave vectors in the star of ${\bf X}$, whose
matrices are given in Table \ref{IRREP}. The allowed basis functions
are given in Table \ref{basis} and are represented in Fig. \ref{X2+}.
Once the configuration for one wave vector is determined, that of the other
%\marginpar{\bf X}
wave vector follows from a $90^{\rm o}$ rotation about the $z$-axis.
So these are two different configurations (one for each $X$ wave vector)
which have the same free energy. These two configurations differ in
their stacking (which takes point ${\bf O}$ into point ${\bf O}'$).
This stacking degeneracy means that this transition takes the I4/mmm
structure into one of two different settings of the space group which
we identify below as Cmca.  One sees that in this distortion the oxygen
octahedra are rotated about the crystal $c$ axis as if they were
interlocking gears.  Notice that in the left panel of Fig. \ref{X2+} 
all the clockwise turning octahedra have $r_x > r_y$ and all the 
counterclockwise ones have $r_x<r_y$,
whereas in the right panel the clockwise turning octahedra have $r_y>r_x$
and the counterclockwise ones have $r_y<r_x$, where $r_x$ is the radius of the
elliptically distorted octahedra along $x$ (or nearly $x$) and $r_y$ is the
radius along $y$.  Symmetry does not fix the sign of the radial distortion.
Changing the sign of the radial distortion would lead to a different
(inequivalent)
structure in which the sign of all the radial arrows for both wave vectors
would be changed.  It would be interesting to observe this radial distortion
in either (or both) calculations and experiment.

As before, we introduce OP's $Q_2({\bf q}_1)$ and $Q_2({\bf q}_2)$
by considering the free energy for which the distortion from tetragonal
is given by $Q_2({\bf q}_1)$ times the distortion for ${\bf q_1}$ plus
$Q_2({\bf q}_2)$ times the distortion for ${\bf q}_2$.  
As in Eq. (\ref{TRANSEQ}), the transformation properties of these OP's
are determined by the matrices of Table \ref{IRREP}.  As before,
wave vector conservation requires that the free energy be a function of
$Q_2({\bf q}_1)^2$ and $Q_2({\bf q}_2)^2$.  Using the transformation properties
of the OP's, we find that
\begin{eqnarray}
{\cal F} ({\bf X}_2) &=& \frac{a'}{2} (T-T_2) [Q_2({\bf q}_1)^2 
+ Q_2({\bf q}_2)^2 ] \nonumber \\ && \
+ \frac{1}{4} u' [ Q_2({\bf q}_1)^2  + Q_2({\bf q}_2)^2 ]^2 \nonumber \\ 
&& \ +  v' [Q_2({\bf q}_1) Q_2({\bf q}_2) ]^2 + {\cal O} (Q^6) \ ,
\nonumber \end{eqnarray}
where $a'>0$, $u'>0$, and $T_2$ is the temperature at which this irrep
%\marginpar{\bf XII}
would become active, if it were the only relevant irrep.  
%\marginpar{\bf XIII}
The free energy of the coupling to strains is
\begin{eqnarray}
V_{Q,\epsilon} &=& \alpha' \epsilon_{xy} [ Q_2({\bf q}_1)^2 
- Q_2({\bf q}_2)^2] + [\beta' (\epsilon_{xx}+\epsilon_{yy})
\nonumber \\ && \
+ \gamma' \epsilon_{zz}] [Q_2({\bf q}_1)^2+ Q_2({\bf q}_2)^2] \ ,
\nonumber \end{eqnarray}
where $\alpha'$, $\beta'$, and $\gamma'$ are arbitrary coefficients.
The coupling to the shear strain is allowed
%\marginpar{\bf XIII}
because both $\epsilon_{yx}$ and $Q_2({\bf q}_1)^2-Q_2({\bf q}_2)^2$ are
odd under ${\cal R}_4$ and even under $m_d$.
The term $[\epsilon_{xx}-\epsilon_{yy}] [Q_2({\bf q}_1)^2-Q_2({\bf q}_2)^2]$
is not allowed because it is not invariant under $m_d$.  Since this
structure is even under inversion it can not have a spontaneous polarization,
so $F_{Q,{\bf P}}$, the interaction linear in the polarization is zero.

Now we discuss the phases which Landau theory predicts. If $v'>0$, then
either $Q_2({\bf q}_1)=0$ or $Q_2({\bf q}_2)=0$, so that
$\epsilon_{xy} \not= 0$ and a detailed analysis shows that the resulting
structure is Cmca (\#64).  Similarly, if $v'<0$, then
one has a double-${\bf q}$ state with
$|Q_2({\bf q}_1)| = |Q_2({\bf q}_2)|$, the orthorhombic distortion vanishes,
and we have the structure P4/mbm (\#127). The
other possibility listed in Ref. \onlinecite{UTAH} is Pbnm (\#35).
As before, we ignore this possibility because
it requires the accidental vanishing of $v'$.  The first principles
calculations of BF indicate that $v'>0$  as required by Fig. \ref{FIG2}.

\subsection{$\Gamma_5^-$}

The allowed distortion of the two-dimensional irrep $\Gamma_5^-$ is given by
the basis functions of Table \ref{basis} and this distortion breaks inversion 
symmetry.  As before, we assume that the distortion $\Psi$ from
I4/mmm is given by $Q_{5,1}$ times the distortion $\Psi_1^{(5)}$ plus
$Q_{5,2}$ times the distortion for $\Psi_2^{(5)}$, where the $\Psi$'s
are given in Table \ref{basis} and are seen to transform simlarly to $x$
or $y$.  It follows that the OP's $Q_{5,1}$ and $Q_{5,2}$, transform as $x$
and $y$, respectively.  So under tetragonal symmetry the expansion of the
free energy $F$ of the distortion $\Psi$ in powers of $Q_{5,1}$ and $Q_{5,2}$
assumes the form
\begin{eqnarray}
{\cal F}(\Gamma_5) &=& \frac{a''}{2} (T-T_5) (Q_{5,1}^2 + Q_{5,2}^2) 
+ \frac{1}{4} u''(Q_{5,1}^2+Q_{5,2}^2)^2
\nonumber \\ && \
+ v'' Q_{5,1}^2Q_{5,2}^2
+ {\cal O} (Q^6) \ ,
\label{F5EQ} \end{eqnarray}
where $a''>0$, $u''>0$, and $T_5$ is the temperature at which irrep
$\Gamma_5$ becomes active (if it were the only  relevant irrep). Since
BF find that $\Gamma_5^-$ is not actually unstable for CMO, $T_5<0$.
Using the fact (see Tables \ref{basis} or \ref{IRREP}) that $Q_{5,1}$ and
$Q_{5,2}$ transform like $x$ and $y$, we see that
\begin{eqnarray}
V_{Q\epsilon} &=& \alpha \epsilon_{zz} (Q_{5,1}^2 + Q_{5,2}^2)
+ \beta (\epsilon_{xx} + \epsilon_{yy}) (Q_{5,1}^2 + Q_{5,2}^2)
\nonumber \\ && \
+ \gamma (\epsilon_{xx}-\epsilon_{yy})(Q_{5,1}^2 -Q_{5,2}^2)
+ \delta \epsilon_{xy}Q_{5,1}Q_{5,2} \ .
\label{F5Qe} \end{eqnarray}
Because this phase is not centrosymmetric, it can support a nonzero
spontaneous polarization.  This will be discussed in a later subsection.

The free energy of Eq. (\ref{F5EQ}) gives rise to
two principal scenarios. If $v''>0$, then either $Q_{5,1}$ or
$Q_{5,2}$ (but not both) condense at $T_5$. Then Eq. (\ref{F5Qe}) indicates
that $\epsilon_{xy}=0$ and $\epsilon_{xx} \not= \epsilon_{yy}$ and
a detailed analysis of the distortions indicates that we condense into
space group Imm2 (\#44).  
Alternatively, if $v''<0$ then Eq. (\ref{F5EQ}) indicates that
$|Q_{5,1}|=|Q_{5,2}|$.  Equation (\ref{F5Qe}) implies that
$\epsilon_{xy} \not= 0$ and $\epsilon_{xx} = \epsilon_{yy}$ and
we condense into space group F2mm (\#42).
As before, these distorted space groups agree with 
the results in Ref. \onlinecite{UTAH}. However, an additional
subgroup is listed as realizable from this irrep, namely
Cm (\#8).  As before, to realize this possibility
requires the accidental vanishing of $v''$, a possibility we reject.
The first principles calculations of BF imply that $v''<0$ so that
space group F2mm would be realized if $\Gamma_5^-$ were to condense first
and both OP's would have equal magnitude.  There are then four possible
domains of ordering corresponding to independently choosing the signs
of $Q_{5,1}$ and $Q_{5,2}$.  

\section{COMBINING IRREPS}

\begin{figure}[h!]
\begin{center}
\includegraphics[width=14.0 cm]{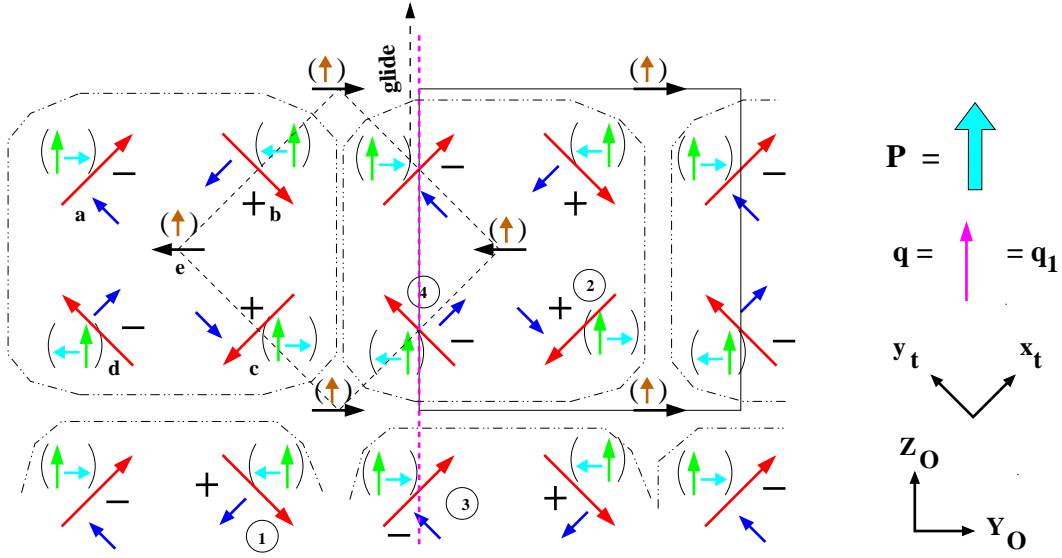}
\caption{\label{FIG6} (Color online)
The symmetry of the displacements of the equatorial oxygen ions 
at tetragonal sites $(\pm 1/2,0,\tau)$ and $(0,\pm 1/2, \tau)$ and
that of apical oxygens ({\it e. g.} e) at $(0,0,\chi)$ in the Cmc2$_1$ phase.
Arrows represent displacements in the $x_t$-$y_t$ plane and the $+$ and
$-$ signs those in the $z_t$ direction.  Each octahedron
({\it e. g.} a, b, c, d, e) is surrounded by a dash-dot oval.  The tetragonal
(orthorhombic) unit cell is bounded by the dashed (full) rectangle.
The spontaneous polarization ${\bf P}$ and the wave vector ${\bf q}$ are shown
at the right along with the tetragonal and orthorhombic coordinate axes.
The displacements of irrep $\Gamma_5^-$ (which contribute to 
${\bf P}\not= 0$) are in parentheses. The other symbols have
the following meaning.  The octahedral rotation (the largest arrows) about
$z_t$ with the smallest arrows for the radial distortion come from irrep
$X_2^+$ and that about $Z_O$ (the apical arrows, the +'s, and the -'s)
comes from irrep $X_3^-$.  The only symmetries remaining in the
Cmc2$_1$ phase are $m_z\equiv m_X=+1$ and a glide plane whose
mirror is the dashed line and whose displacement is the dashed arrow. 
The glide operation takes site 1 into 2 and 3 into 4.}
\end{center}
\end{figure}

Now we consider what happens when we condense a second irrep.  Although we
assume that $X_3^-$ is the first irrep to condense, our discussion could be
framed more generally when the ordering of condensation of the irreps
is arbitrary.  We assume a quadratic free energy ${\cal F}_2$ of the form
\begin{eqnarray}
{\cal F}_2 &=& \frac{q}{2} (T-T_3) [Q_3({\bf q}_1)^2+Q_3({\bf q}_2)^2]
\nonumber \\ && \
+ \frac{a'}{2} (T-T_2) [Q_2({\bf q}_1)^2 + Q_2({\bf q}_2)^2]
\nonumber \\ && \ + \frac{a''}{2} (T-T_5) (Y_5^2+Z_5^2) \ ,
\nonumber \end{eqnarray}
where $Y_5=X_{5,1}-X_{5,2}$, $Z_5=X_{5,1}+X_{5,2}$, we assume that 
$T_3 > T_2 > T_5$.  For the RP systems wave vector conservation and inversion
invariance implies that the cubic terms in the free energy must be of the form
\begin{eqnarray}
V_C &=& Q_2({\bf q}_1) Q_3({\bf q}_1) [rQ_{5,1} + t Q_{5,2}]
\nonumber \\ && \
+ Q_2({\bf q}_2) Q_3({\bf q}_2) [sQ_{5,1} + uQ_{5,2}] \ .
\nonumber \end{eqnarray}
To make this invariant under $m_d$ we require $r=t$ and $s=-u$.  Invariance
under ${\cal R}_4$ leads to $s=r$, so that we may write
\begin{eqnarray}
V_C &=& -r[Q_2({\bf q}_1) Q_3({\bf q}_1) Z_5
+ Q_2({\bf q}_2) Q_3({\bf q}_2) Y_5 ] \ .
\label{CUBIC} \end{eqnarray}
%\marginpar{\bf XVII}
As the temperature is reduced, $Q_3$ is the first OP to condense, with
either $Q_3({\bf q}_1) \not= 0$ or $Q_3({\bf q}_2) \not= 0$, as dictated
by the quartic terms we considered previously in Eq. (\ref{F3-}).
Then, assuming $Q_3({\bf q}_1)$ has condensed, we have effectively
\begin{eqnarray}
{\cal F}_3 &=& w \langle Q_3({\bf q}_1) \rangle Q_2({\bf q}_1)  Z_5 \equiv
w' Q_2({\bf q}_1)  Z_5  \ ,
\nonumber \end{eqnarray}
where $\langle Q_3({\bf q}_1) \rangle$ indicates the value of $Q_3({\bf q}_1)$
which minimizes the free energy.
Thus the $Q_2$ and $Q_5$ variables are governed by the quadratic free energy
\begin{eqnarray}
{\cal F}_{Q_2,Z} &=& \frac{1}{2} (T-T_2) Q_2({\bf q}_1)^2
+ \frac{1}{2} (T-T_5) Z_5^2 \nonumber \\ && \
+ w \langle Q_3({\bf q}_1)\rangle  Q_2({\bf q}_1) Z_5
+ \frac{1}{2} (T-T_2)Q_2({\bf q}_2)^2 
\nonumber \\ && \ + \frac{1}{2} (T-T_5) Y_5^2
+ w \langle Q_3({\bf q}_2)\rangle Q_2({\bf q}_2) Y_5 \ .
\nonumber  \end{eqnarray}
Suppose that $Q_3({\bf q}_1)$ is the first variable to condense.
The effect of the cubic term is to couple $Q_2({\bf q}_1)$ and $Z_5$ so that
the variable that next condenses as the temperature is
lowered is a linear combination of $Q_2({\bf q}_1)$ and $Z_5$.
If $w'$ is small compared to $T_2-T_5$, then the new transition
temperature will be approximately $\tilde{T_2} \approx T_2+ w'/(T_2-T_5)$
and the condensing variable ${\tilde Q}_2({\bf q}_1)$ will dominantly be
$Q_2({\bf q}_1)$ with a small amount of $Z_5$ admixed to it.
The important conclusion is that we have two families of OP's:
$(Q_3({\bf q}_1),Q_2({\bf q}_1),Z_5)$ and $(Q_3({\bf q}_2),Q_2({\bf q}_2),Y_5)$.
At the highest transition one OP (we assume it to be $Q_3$) of one of the two
families will condense.  At a lower temperature the two other OP's of that
family will condense. Independently of which OP first condenses one will
reach one of the equivalent domains of the same final state, as Fig.
\ref{FIG2} indicates.  There are eight[\onlinecite{EIGHT}]
such equivalent domains because we can independently choose between
a) the wave vectors ${\bf q}_1$ and ${\bf q}_2$, b) the sign of $Q_3({\bf q})$,
and c) the sign of ${\tilde Q}_2({\bf q})$.  These domains are discussed
in detail in Sec. VI. For simplicity, we do not extend 
this analysis to include several copies of the various irreps involved.
The symmetry of the displacements when the irreps of the ${\bf q}_1$
family are present is shown in Fig. \ref{FIG6}.

\vspace{0.2 in} \noindent
The cubic term of Eq. (19) guarantees that when $Q_3({\bf q}_n)$
condenses, the lower-temperature transition always
involves the condensation of $Q_2({\bf q}_n)$ and the
appropriate $\Gamma_5^-$ OP.  We selected this cubic
interaction to be dominant in order to arrive finally at the observed
Cmc2$_1$ phase, as discussed in Appendix \ref{APPA}.
There is also a cubic interaction of the form
\begin{eqnarray}
V_C^\prime &=& r[ Q_2({\bf q}_1) Q_3({\bf q}_2) V({\bf q}_M)
+ Q_2({\bf q}_2) Q_3({\bf q}_1) W({\bf q}_M)] \ ,
\nonumber \end{eqnarray}
where ${\bf q}_M = (1,0,0)$ and $V_M$ and $W_M$ are operators
that transform according to the appropriate two-dimensional 
irrep ($M_5^-$) so that $V_C^\prime$ is an invariant.  This
interaction (in contrast to $V_C$) involves
a doubling of the size of the unit cell at ${\tilde T}_2$. Since no such
doubling has been seen, we conclude that $V_C^\prime$ 
is dominated by $V_C$.

\begin{figure}[h!]
\begin{center}
\includegraphics[width=6.0 cm]{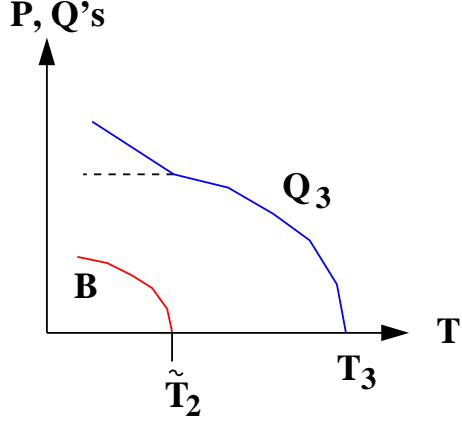}
\caption{\label{QP} (Color online) Temperature dependence of $Q_3$
for $T$ near $T_3$ and (lower curve) of $Q_3$, $Q_2$, and
$Q_5$ for $T$ near $\tilde {T_2}$, as discussed below Eq. (\ref{QQ})
and of ${\bf P}$ as described by Eq. (\ref{EQ30}).}
\end{center}
\end{figure}

It is interesting to consider the mean-field temperature dependence of
$Q_3({\bf q}_1)$, for example, from the free energy
\begin{eqnarray}
{\cal F} &=& \frac{a}{2} (T-T_3) Q_3({\bf q}_1)^2 + \frac{1}{4} u 
Q_3({\bf q}_1)^4 \nonumber \\ && \
+ w Q_3({\bf q}_1) Q_2({\bf q}_1) [Q_{5,1}+Q_{5,2}] \ .
\nonumber \end{eqnarray}
For $T$ slightly below $T_3$ one has $|Q_3({\bf q}_1)| \sim (T_3-T)^{1/2}$.
For $T$ near $\tilde {T_2}$ we treat the term in $w$ perturbatively and find
for $T<T_2$ that[\onlinecite{FN3}]
\begin{eqnarray}
Q_3({\bf q}_1) & \approx& [(a/u)(T_3-T)]^{1/2} \nonumber \\ && \
+ \frac{wQ_2({\bf q}_1)(Q_{5,1}+Q_{5,2})}{2a(T_3-T)} \ .
\label{QQ} \end{eqnarray}
Slightly below $\tilde {T_2}$
the variables $Q_2({\bf q}_1)$ and $(Q_{5,1}+Q_{5,2})$
are proportional (with different constants of proportionality) to
$(\tilde {T_2}-T)^{1/2}$. Thus the second term in Eq. (\ref{QQ}),
which is only nonzero for $T<\tilde {T_2}$, is proportional to
$(\tilde {T_2}-T)$. These results are illustrated in Fig. \ref{QP}.

One can similarly analyze the temperature dependence of the strains.
From Eq. (\ref{G3STRAIN}) we see that the strain $\epsilon_{xy}$
(in tetragonal coordinates) is zero for $T>T_3$.  Just below $T_3$
one has
\begin{eqnarray}
\epsilon_{xy} = \alpha [Q_3^-({\bf q}_2)^2 - Q_3^-({\bf q}_1)^2]
/ c_{44}\ .
\nonumber \end{eqnarray}
Since $Q_3 \propto (T_3-T)^{1/2}$, this indicates that just below
$T_3$ one has
\begin{eqnarray}
\epsilon_{xy} = \pm (T_3-T) \ ,
\nonumber \end{eqnarray}
the sign depending on which $Q_3({\bf q}_n)$ has condensed.
In addition, the coupling between strains and the orientational
OP's indicate that at the structural transitions there will be a
jump in slope of the diagonal strains $\epsilon_{\alpha \alpha}$.

\section{Coupling to Dielectric and Magnetic Order}

In the following two subsections we consider the coupling between the
lattice distortions and a) the spontaneous polarization
and b) magnetic long range order.

\subsection{Dielectric Coupling}

The dielectric free energy is
\begin{eqnarray}
{\cal F}_{\rm D} &=& \frac{1}{2} \chi_E^{-1} [ \vec P ]^2 + V_{\rm D} \ ,
\nonumber \end{eqnarray}
where $V_{\rm D}$ is the coupling with the distortion modes and which is
linear in the spontaneous polarization ${\bf P}$.  This coupling will be
zero for a structure which has a center of inversion symmetry. 
$\Gamma_5^-$ is the only irrep which, by itself, breaks inversion and for
it we have
\begin{eqnarray}
V_{\rm D}^{(5)} &=& - \lambda [ Q_{5,1} P_x + Q_{5,2} P_y ] \ .
\nonumber \end{eqnarray}
This follows because $Q_{5,1}$ transforms like $x$ and $Q_{5,2}$ transforms
like $y$. Using this transformation property, we infer from Eq. 
(\ref{CUBIC}) a contribution to $V_{\rm D}$ of the form
\begin{eqnarray}
V_{\rm D}^{(2,3)} &=& - \lambda' [ 
Q_2({\bf q}_1) Q_3({\bf q}_1)(P_x+P_y) \nonumber \\ && \
+ Q_2({\bf q}_2) Q_3({\bf q}_2)(P_x-P_y)] \ .
\nonumber \end{eqnarray}
These couplings indicate that
\begin{eqnarray}
P_x &=& \lambda \chi_E Q_{5,1} + \lambda' \chi_E 
[Q_2({\bf q}_1) Q_3({\bf q}_1) + Q_2({\bf q}_2) Q_3({\bf q}_2)] \nonumber \\
P_y &=& \lambda \chi_E Q_{5,2} + \lambda' \chi_E 
[Q_2({\bf q}_1) Q_3({\bf q}_1) - Q_2({\bf q}_2) Q_3({\bf q}_2)] \ .
\nonumber \\ \label{PQEQ} \end{eqnarray}
We may simplify this by minimizing the free energy of Eqs. (\ref{F5EQ})
and (\ref{CUBIC}) to write
\begin{eqnarray}
Q_{5,1} &=& \frac{r}{T-T_5} \Biggl( 
Q_2({\bf q}_1) Q_3 ({\bf q}_1) + Q_2({\bf q}_2) Q_3 ({\bf q}_2) \Biggr)
\nonumber \\
Q_{5,2} &=& \frac{r}{T-T_5} \Biggl(
 Q_2({\bf q}_1) Q_3 ({\bf q}_1) - Q_2({\bf q}_2) Q_3 ({\bf q}_2) \Biggr) \ .
\nonumber \end{eqnarray}
Thus
\begin{eqnarray}
P_x &=& \tau \Biggl( Q_2({\bf q}_1) Q_3 ({\bf q}_1)
+ Q_2({\bf q}_2) Q_3 ({\bf q}_2) \Biggr) \nonumber \\
P_y &=& \tau \Biggl( Q_2({\bf q}_1) Q_3 ({\bf q}_1)
- Q_2({\bf q}_2) Q_3 ({\bf q}_2) \Biggr) \ ,
\label{EQ30} \end{eqnarray}
%\marginpar{\bf XVIII}
where $\tau= \chi_E [ \lambda' + \lambda r /(T-T_5)]$.  As shown in Appendix 
\ref{APPC}, the dielectric constant will have a small amplitude divergence 
at a temperature $T={\tilde T}_2$ near $T_2$, as is often seen in systems
with magnetization induced polarization.[\onlinecite{R1,R2}].
Note that there are two mechanisms for a spontaneous polarization
proportional respectively to $\lambda'$ and $\lambda$.  The term in $\lambda$
may be viewed as being the polarization due to displacement of the charged
ions in the polar irrep $\Gamma_5^-$. This displacement is induced by the
presence of the other two OP's $Q_2$ and $Q_3$.  The term in $\lambda'$
is due to  the modification in the electronic structure proportional to
$Q_2({\bf q}) Q_3({\bf q})$ when the displacement due to
$\Gamma_5^-$ is zero.  The numerical work
of BF indicates that there is a large ($\sim 5 \mu {\rm C/cm}^2$)
spontaneous polarization which arises when $Q_5$ is zero and
therefore indicates that mechanism b) is dominant.

For $T$ slightly below $\tilde {T_2}$, $Q_3$ is essentially  constant and
the other $Q$'s are proportional to $(\tilde {T_2}-T)^{1/2}$ within
mean field theory, as illustrated in Fig. \ref{QP}.  Equation
(\ref{EQ30}) indicates that ${\bf P}$ is parallel to ${\bf q}$  and
is proportional to $Q_2Q_3$, so that near ${\tilde T}_2$ one has the mean-field
result that $|{\bf P}| \sim (\tilde T_2 -T)^{1/2}$, as shown in Fig.
\ref{QP}.  From the work of Mostovoy,[\onlinecite{MOST1}] magnetically
induced polarization was not expected to have ${\bf P} \parallel {\bf q}$,
although ${\bf P} \parallel {\bf q}$, was found 
experimentally[\onlinecite{RBFE}] and explained from general symmetry
arguments.[\onlinecite{RBFE,TAK}] Equation (\ref{PQEQ}) indicates that
a polarization will be induced parallel to ${\bf q}$
along one of the four $(1,1,0)$ directions
according to the signs of the OP's as indicated by Eq. (\ref{PQEQ}). 

\subsection{Magnetism}

\begin{figure}[ht]
\begin{center}
\includegraphics[width=8.6 cm]{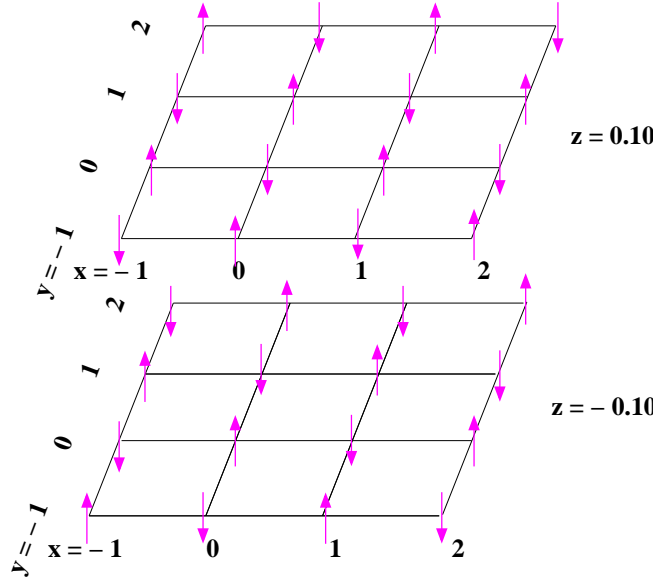}
\caption{\label{AF} (Color online)
A single bilayer of Mn ions when the spins are assumed
to lie in the plane.  This is a $G_X$ configuration.  The stacking
of adjacent bilayers (not shown) is such that they are displaced transversely
by $(1/2,1/2,0)$ relative to one another. The phase of antiferromagnetic
order in adjacent bilayers is determined by the wave vector which
is either ${\bf q}_1$ or ${\bf q}_2$. This configuration of pseudovectors
is odd under inversion about the origin ($x=y=z=0$).}
\end{center}
\end{figure}

\begin{figure}[h!]
\begin{center}
\includegraphics[width=8.6 cm]{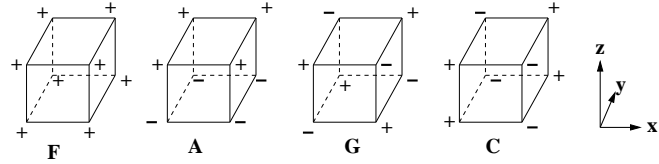}
\caption{\label{W-K} The A, C, F, and G configurations of spin components
for a single bilayer in the Wollan-Koehler[\onlinecite{WK}] scheme.}
\end{center}
\end{figure}

We now discuss the magnetic structures that can appear in this system.
Ca$_3$Mn$_2$O$_7$ becomes antiferromagnetic at $T_N=115$K.[\onlinecite{LOB}]
For this discussion we will work relative to the parent tetragonal lattice
but will introduce the orthorhombic coordinates for the spin
vectors so that
\begin{eqnarray}
(S_X)_o &=& (S_z)_t \ , \hspace{0.2 in}
(S_Y)_o = (S_x)_t + (S_y)_t \ , \hspace{0.2 in} \nonumber \\
&& \ (S_Z)_o = (S_x)_t - (S_y)_t \  . \nonumber
\end{eqnarray}
In the
simplest approximation the antiferromagnetic structure of a single biliayer
(consisting of one layer of Mn ions at $z_t = x_o=\xi$ and another 
at $z_t=x_o=-\xi$) is that of two square lattice antiferromagnets stacked
directly on top of one another [sites at $(0,0,\xi)$ and $(0,0,-\xi)$],
so that all near neighbor interactions proceed via nearly 180$^{\rm o}$
antiferromagnetic Mn-O-Mn bonds.  The first principles calculations
of BF and the data from Ref. \onlinecite{LOB}
indicate that dominantly the spins are  perpendicular to the plane.
This structure is shown in Fig. \ref{AF}.
Therefore we assume that the dominant order parameter is ${\bf G}({\bf q})$,
where we use the Wollan-Koehler[\onlinecite{WK}] symbols to represent
the configurations of a single bilayer shown in Fig. \ref{W-K}.  In the
$G_z \equiv G_X$ configuration one can have either
${\bf q}={\bf q}_1$ or ${\bf q}={\bf q}_2$.  The difference between
$G_X({\bf q}_1)$ and $G_X({\bf q}_2)$ is in the different phase of one
bilayer relative to the other.  In any case all 5 magnetic
nearest neighbors of a central spin are oriented antiparallel to it.

\begin{figure}[h!]
\vspace{0.1 in}
\begin{center}
\includegraphics[width=8.6 cm]{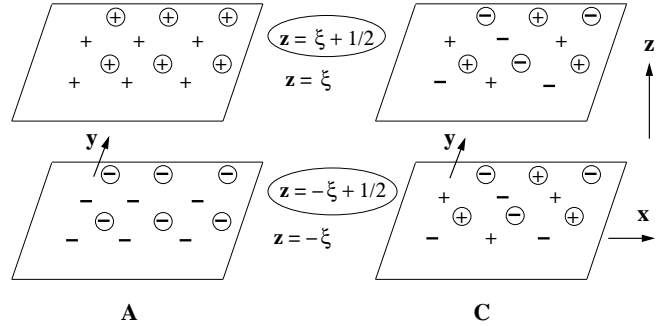}
\caption{\label{ACEG} Spin states of a bilayer of Mn ions.  The plus and
minus signs represent the signs of any component of spin.  Left: the ``A"
configuration.  Right: the ``C" configuration. The circled symbols represent
the spins in planes at $z=1/2+\xi$ and $z=1/2-\xi$ for ${\bf q}={\bf q}_1$. 
For ${\bf q}={\bf q}_2$ the circled $+$ and circled $-$ signs of the
C configuration are interchanged.}
\end{center}
\end{figure}

We will accommodate the following structures based on the parent tetragonal
lattice.  We have choices for the wave vector, namely ${\bf q}=0$,
${\bf q}={\bf q}_1$ and ${\bf q}_2$ and $S_\alpha (0,0,\xi)= \pm
S_\alpha (0,0,-\xi)$.  So for each component of spin we have six
candidate structures. If ${\bf q}=0$ and $S_\alpha (0,0,-\xi)=S(0,0,\xi)$,
then we have the ``F'' (ferromagnetic) structure.
If ${\bf q}=0$ and $S_\alpha (0,0,-\xi)=-S(0,0,\xi)$, then
we have the ``A'' structure shown in Fig. \ref{W-K}.
If ${\bf q}={\bf q}_n$, then each plane of the bilayer consists of a
square lattice antiferromagnet. The two planes of the bilayer can be
coupled either so that adjacent spins in the planes are parallel (this is
the ``C'' structure) or so that they are antiparallel (this is the ``G"
structure), as shown in Fig. \ref{W-K}.
The C and G structures each come in two versions depending on 
whether ${\bf q}={\bf q}_1=(1/2,1/2,0)$ or ${\bf q}={\bf q}_2=(1/2,-1/2,0)$,
as discussed in the caption to Fig. \ref{ACEG}.  We now introduce
orthorhombic coordinates, so that $S_X=S_z$,
$S_Y = S_x - S_y$, and $S_Z= S_x + S_y$.  Then we have the symmetries
given in Table \ref{SYM}.

\begin{table} [h!]
\caption{\label{SYM} The symmetry for components $X$, $Y$, and $Z$ of the
spin, a pseudovector.  Here $S_X=S_z$, $S_Y=S_x+S_y$, $S_Z=S_x-S_y$, where
capitals refer to orthorhombic and lower case to tetragonal.}
\vspace{0.2 in}
\begin{tabular} {||c c| c c c |c c c |c c c ||}
\hline \hline
Structure & \ \ ${\bf q}$\ \ & \multicolumn{3} {c|} {${\cal I}$} &
\multicolumn{3} {c|} {$m_z$} & \multicolumn{3} {c||} {$m_d$} \\
\hline
&& $Y$ & $Z$ & $X$ & $Y$ & $Z$ & $X$ & $Y$ & $Z$ & $X$ \\
\hline
F & $0$         & $+$ &$+$ &$+$ &  $-$ & $-$ & $+$ & $+$ & $-$ & $-$   \\
A & $0$         & $-$ &$-$ &$-$ &  $+$ & $+$ & $-$ & $+$ & $-$ & $-$   \\
G & ${\bf q}_1$ & $-$ &$-$ &$-$ &  $+$ & $+$ & $-$ & $+$ & $-$ & $-$   \\
G & ${\bf q}_2$ & $-$ &$-$ &$-$ &  $+$ & $+$ & $-$ & $+$ & $-$ & $-$   \\
C & ${\bf q}_1$ & $+$ &$+$ &$+$ &  $-$ & $-$ & $+$ & $+$ & $-$ & $-$   \\
C & ${\bf q}_2$ & $+$ &$+$ &$+$ &  $-$ & $-$ & $+$ & $+$ & $-$ & $-$   \\
\hline \hline
\end{tabular}
\end{table}

The magnetic free energy is $F = F_G +V$, where $F_G$ is the free energy
of the G structure:
\begin{eqnarray}
F_G &=& \frac{1}{2} \sum_\alpha \Biggl[ (T-T_N+K_\alpha) \sum_n G_\alpha
({\bf q}_n)^2 \nonumber \\ && \
- \alpha' \sum_n G_\alpha ({\bf q}_n)^2 Q_3({\bf q}_n)^2
\Biggr]  + {\cal O} [ G({\bf q}_n)^4 ]
\nonumber \end{eqnarray}
and
\begin{eqnarray}
V &=& \frac{1}{2} \sum_\alpha \Biggl[ \mu_\alpha F_\alpha^2
+ \nu_\alpha C_\alpha({\bf q}_1)^2 + \tau_\alpha A_\alpha^2 \Biggr] \ . 
\label{FQUAD} \end{eqnarray}
The quartic term in $F_G$ is such as to ensure that the magnetic
ordering vector is the same as that of the paramagnetic structure.
Now we want to see what other magnetic OP's are induced by the condensation
of the dominant $G$ ordering in the presence of the nontetragonal distortions.
The magnetoelastic interaction we invoke has to be quadratic in the magnetic
variables in order to be time-reversal invariant.  So we consider a 
cubic potential which contains terms of the form
\begin{eqnarray}
G_X({\bf q}_n) K_\alpha Q_\beta \ ,
\nonumber \end{eqnarray}
where $K$ is $C({\bf q}_n)$, $F$, or $A$ and $Q$ is $Q_2^+({\bf q}_n)$,
$Q_3^-({\bf q}_n)$, $Z_5 \equiv Q_{5,1}+Q_{5,2}$, or
$Y_5 \equiv Q_{5,1}-Q_{5,2}$. We start by considering only terms involving
${\bf q}_1$.  The terms involving ${\bf q}_2$ will later be obtained from
those involving ${\bf q}_1$ by applying the four-fold rotation ${\cal R}_4$.
The only terms which are consistent with inversion symmetry and wave vector
conservation are those of the form
\begin{eqnarray}
&& G_X({\bf q}_1) A_\alpha Q_2^+({\bf q}_1) \ , \ \ \
G_X({\bf q}_1) F_\alpha Q_3^-({\bf q}_1) \ , \nonumber \\ && \
G_X({\bf q}_1) C_\alpha ({\bf q}_1) Z_5 \ .
\nonumber \end{eqnarray}
($Y_5=0$ for ${\bf q}={\bf q}_1$.) We now use Table \ref{SYM} 
to require invariance under $m_d$ and $m_z$, so that the
interaction which has the correct symmetry is
\begin{eqnarray}
aG_X({\bf q}_1) F_Y Q_3^-({\bf q}_1) +
b G_X({\bf q}_1) C_Z({\bf q}_1) Z_5 \ .
\label{aaEQ} \end{eqnarray}
Now we use
\begin{eqnarray}
{\cal R}_4 G_X ({\bf q}_n) &=& G_X ({\bf q}_{3-n})\ , \ \ \
{\cal R}_4 F_Y = -F_Z \ , \nonumber \\
{\cal R}_4 C_Z ({\bf q}_1) &=& -C_Y ({\bf q}_2) \ , \ \ \
{\cal R}_4 Q_3^-({\bf q}_1) = Q_3^-({\bf q}_2) \ , \
\nonumber \\ {\cal R}_4 Z_5 &=& Y_5 \ .
\nonumber \end{eqnarray}
Thus, in all, the lowest order magnetoelastic coupling $V_{MQ}$ is
\begin{eqnarray}
V_{MQ} &=& aG_X({\bf q}_1) F_Y Q_3^-({\bf q}_1) +
b G_X({\bf q}_1) C_Z({\bf q}_1) Z_5 \nonumber \\
&& - aG_X({\bf q}_2) F_Z Q_3^-({\bf q}_2)
- b G_X({\bf q}_2) C_Y({\bf q}_2) Y_5 \ .
\nonumber \\ \label{EQMQ} \end{eqnarray}
To see what this means, it is helpful to recall that ${\bf q}_1$
(${\bf q}_2$) lies along the orthorhombic $Z$ ($Y$) direction.
Thus the weak ferromagnetic moment $F$ is perpendicular to ${\bf q}$.
Note that the wave vector is already selected as soon as tetragonal
symmetry is broken. Say ${\bf q}_1$ is selected. Then, in addition,
the sign of $Q_3^-({\bf q}_1)$ was also selected when tetragonal symmetry
was broken.  Then, when magnetic long-range order appears, it can have
either sign of $G_X({\bf q}_1)$, but the sign of $G_X({\bf q}_1)F_Z$
is fixed by the interactions within the system.  

\begin{figure}[h!]
\begin{center}
\includegraphics[width=8.6 cm]{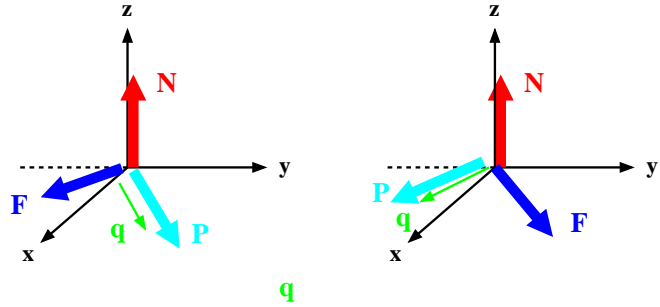}
\caption{\label{FIGNPM} (Color online) The equilibrium orientation of
the staggered magnetization (${\bf N}$), the polarization (${\bf P}$),
and the weak ferromagnetic moment (${\bf F}$) for ${\bf q}={\bf q}_1$
(left panel) and for ${\bf q}={\bf q}_2$ (right panel). The axes along
which these vectors lie are fixed by symmetry. However, one can find
domains in which any (or all) these vectors can be reversed.
For instance, ${\bf P}$ is proportional to $Q_2Q_3$ whereas
${\bf F}$ is proportional to $Q_3$.}
\end{center}
\end{figure}

Assuming the G configuration to be dominant and using Eq. (\ref{FQUAD}),
we thus have two scenarios.  If ${\bf q}={\bf q}_1$, then we have the
magnetic OP's
\begin{eqnarray}
&& \ [G_X({\bf q}_1), F_Y , C_Z({\bf q}_1)] \ ,
\ \ \ \ \ {\rm with} \nonumber \\
&& \frac{F_Y}{G_X({\bf q}_1)} = - \frac{a Q_3^-({\bf q}_1)}{\mu_Z} \ ,
\hspace{0.2 in}  \frac{C_Z({\bf q}_1)}{G_X({\bf q}_1)}
= - \frac{b Z_5}{\nu_Y} \ .
\label{EQ38A} \end{eqnarray}
If ${\bf q}={\bf q}_2$, then we have the magnetic OP's
\begin{eqnarray}
&& \ [G_X({\bf q}_2), C_Y({\bf q}_2 , F_Z)] \ , \ \ \ \ \ {\rm with} 
\nonumber \\ && \frac{C_Y({\bf q}_2)}{G_X({\bf q}_2)}
= \frac{b Y_5}{\nu_Z} \ , \hspace{0.2 in}
\frac{F_Z}{G_X({\bf q}_2)} = \frac{a Q_3^-({\bf q}_2)}{\mu_Y} \ .
\label{EQ38B} \end{eqnarray}
In all the above results, since the cubic coupling combines $Z_5$
with $Q_3^-({\bf q}_1)Q_2^+({\bf q}_1)$ and similarly for ${\bf q}={\bf q}_2$,
we should replace $Z_5$ by a linear combination of $Z_5$ and
$Q_3^-({\bf q}_1)Q_2^+({\bf q}_1)$ and $Y_5$ by a linear combination
of $Y_5$ and $Q_3^-({\bf q}_2)Q_2^+({\bf q}_2)$. The result of
Eqs. (\ref{EQ38A}) and (\ref{EQ38B}) agrees with the magnetic structure
determination of Ref. \onlinecite{LOB} and with the symmetry analysis of
BF, except that here we emphasize the
relation of the magnetic ordering to the preestablished wave vector.
These results are summarized by Fig. \ref{FIGNPM}.

It is interesting to note the possibility of switching the direction of 
the polarization (magnetization) by application of a sufficiently strong
magnetic (electric) field. For this discussion it is useful ro refer to
Fig. \ref{FIGNPM}.  Suppose the sample initially has condensed wave vector 
${\bf q}_1$.  (See the left panel of Fig. \ref{FIGNPM}.)  Applying a large
enough magnetic field {\it in the $Z$ direction}
({\it i. e.} parallel to ${\bf q}_1$) will cause a reorientation of
the magnetization of the system so that $F_Y=0$ and $F_Z \not= 0$.  Thus
$V_{MQ}$ indicates that the system will make a transition from
wave vector ${\bf q}_1$ to wave vector ${\bf q}_2$. Since the direction
of the polarization is tied to the direction of the wave vector, this
transition will also involve a rotation of the polarization.
Likewise, suppose the sample initially has the polarization and the condensed
wave vector collinear and parallel to ${\bf q}_1$.  Equation (\ref{EQ38A})
indicates that the ferromagnetic moment is along $Y$ (which is
perpendicular to ${\bf q}_1$). Then,
if one applies a large enough electric field parallel to ${\bf q}_2$,
then the polarization will rotate into the direction of ${\bf q}_2$
and since the polarization and wave vector are constrained to be collinear,
the wave vector will now be along ${\bf q}_2$.  Then Eq. (\ref{EQMQ})
indicates that the weak ferromagnetic moment will be rotated into
the $Z$ direction (which is perpendicular to ${\bf q}_2$).
However, the applied fields required to accomplish these switchings
may be extremely large in view of the structural reorganization involved.

Since $F_\alpha$ transforms like a pseudovector, one has symmetry-allowed
interactions with $F_\alpha$ replaced by $H_\alpha$ where ${\bf H}$ is
the applied magnetic field.  So from Eq. (\ref{EQMQ}) we have a
magnetic-field dependent contribution to the free energy of the form
\begin{eqnarray}
V_{\bf H} &=& c_1 G_X({\bf q}_1) Q_3^-({\bf q}_1) H_Y +
c_2 G_X({\bf q}_2) Q_3^-({\bf q}_2) H_Z \ ,
\label{HEQ} \end{eqnarray}
which indicates that when tetragonal symmetry is broken (so that $Q_3$ is
nonzero), the magnetic field acts like a field conjugate to the
antiferromagnetic OP ${G_X}$. Consequently, $\partial G_X/\partial H_\alpha$
will diverge as the lower transition is approached for $\alpha=Y$ or
$\alpha=Z$, according to which wave vector has condensed. This suggests 
a neutron scattering  experiment to measure $G_X$ near the lower
transition as a function of the magnitude and direction of ${\bf H}$.

\section{Domains}

\begin{figure}[h!]
\begin{center}
\includegraphics[width=8.6 cm]{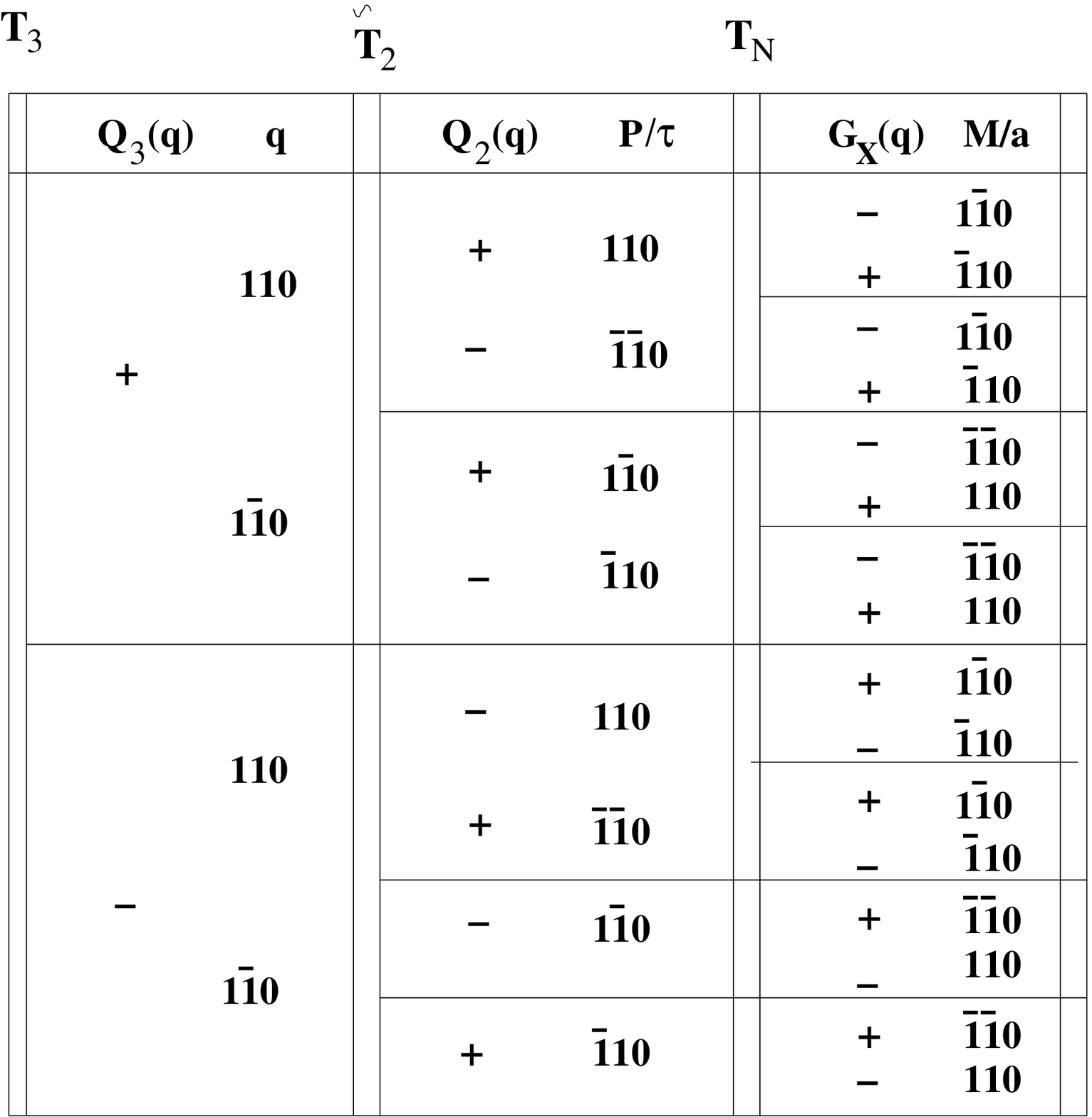}
\caption{\label{DOMFIG} Domains in CMO.  At $T_3$ the value of the wave vector
${\bf q}$ and of $Q_3({\bf q})$ are selected. At each succeeding transition
a two-state order parameter condenses to further break symmetry. The
domains for positive $Q_3({\bf q})$ are macroscopically indistinguishable
from those for negative $Q_3({\bf q})$ because they differ by a unit
translation in the $x$-$y$ plane. The signs of the OP's ${\bf P}$ and
the weak ferromagnetic moment ${\bf M}$ depend on the constants $\tau$
of Eq. (\ref{EQ30}) and $a$ of Eq. (\ref{aaEQ}). For CTO (which is 
nonmagnetic) the section at and below $T_N$ does not apply.}
\end{center}
\end{figure}

Here we discuss in more detail the possible domain structures and
give a brief discussion of the dynamics of domain wall motion.
We first enumerate the various domains of order parameters which can exist
as the temperature is lowered through the various phase transitions.
As a preliminary one should note that within a single domain
the {\it phase} of an OP at wave vector ${\bf q}_k$ can not be experimentally
determined.  However, if more than one such OP is present, then their
{\it relative} phases can be accessed experimentally. In the discussion that
follows we will determine the phase of the OP's relative to that of
$Q_3$ which is not determined. 
As the temperature is lowered through $T_3$, four possible domains
are created, with the choices of sign of $Q_3$ and the two choices
for the wave vector, as shown in Fig. \ref{DOMFIG}.  The value of the wave
vector within a single domain is experimentally accessible via a
scattering experiment.  Although the phase of $Q_3({\bf q})$ can not
be established within a single domain, the fact that different such
domains do exist can be established by observation of a domain wall
separating domains having the same values of the wave vector.  Such an
experiment to observe a so-called phase domain wall has recently been done
in another system.[\onlinecite{BODE}]

Next, as the temperature is lowered through the lower structural transition
at $T={\tilde T}_2$,  the OP $Q_2$ is condensed (and, as we have
seen, is accompanied by $Q_5^-$), giving a total of eight
domains, four with $Q_3>0$ and four with $Q_3<0$.  These sets of four
differ from one another only in the inaccessible phase of the OP $Q_3$.
However, the four domains having a given sign of $Q_3$ can be
distinguished from one another, since they correspond to the
two choices of wave vector (which is easily experimentally accessible)
and the two choices of sign of $Q_2$ which leads to distinct orientations
of the spontaneous polarization (which is also easily experimentally
accessible).

Finally, when the temperature $T_N$ is reached, the OP $G_X({\bf q})$
which describes the antiferromagnetic order is also accessible because
it is coupled to the weak ferromagnetic moment.  In that way we
can identify the 16 domains up to an uncertainty in the sign of
$Q_3$.  As we indicated above, the uncertainty is, in principle,
accessible in that the phase domain wall can be observed.
Note that the transformation $(Q_3, Q_2, G_x) \rightarrow (-Q_3,-Q_2,-G_x)$
leaves the observables ${\bf P}$, ${\bf M}$, and ${\bf q}$ invariant.

\begin{figure}[h!]
\begin{center}
\includegraphics[width=8.6 cm]{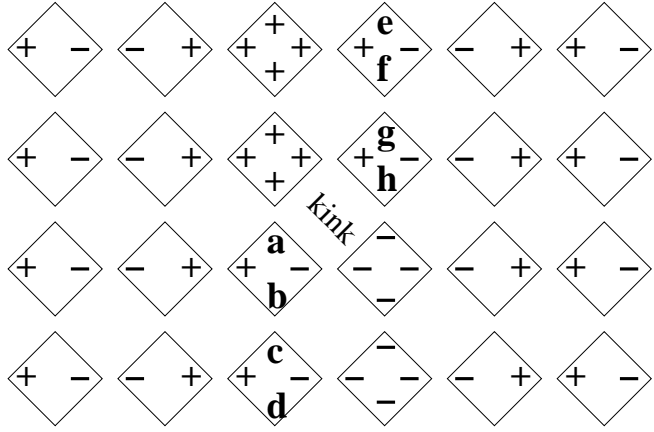}
\caption{\label{DOM1} Displacements of the equatorial oxygens as in Fig.
\ref{FIGX3}.  An antiphase domain wall in the phase of a tilted
octahedral phase in an $n=1$ RP system. The phases to the
left and to the right of the domain wall are indistinguishable
since they only differ in the phase of the ordering.  In this
regard this situation is analogous  to domains in an antiferromagnet.
If the domain wall were a perfect plane, there would be no
distortion energy of the octahedra. However,
when the wall has a kink, as shown, then there are two possibilities.
First, the tilting a, b, c, d can heal to zero, but this distortion
will involve an intraoctahedral distortion of high energy.  Secondly,
it is possible that a, b, c, d (or analogously e, f, g, h) will
alternate in sign, end at an antikink,
so that no intraoctahedral energy is involved, but
the tilting of the octahedra is not the lowest energy tilting of
the established phase and the energy of the kink is 
proportional to the distance between the kink and the antikink.}
\end{center}
\end{figure}

The above discussion assumes the existence of domains which, unlike
ferromagnetic domains, do not have an obvious energetic reason to exist
for $T_3 > T > {\tilde T}_2$, where ${\bf P}$ is zero.
In particular, the phase domains, if they exist, would appear below the
upper ordering temperature $T_3$, where $Q_3({\bf q})$ condenses,
in which one has, for a given wave vector (which can be selected by
applying a suitable shear stress), the two possible choices of
sign of $Q_3({\bf q})$.

\begin{figure}[h!]
\begin{center}
\includegraphics[width=8.6 cm]{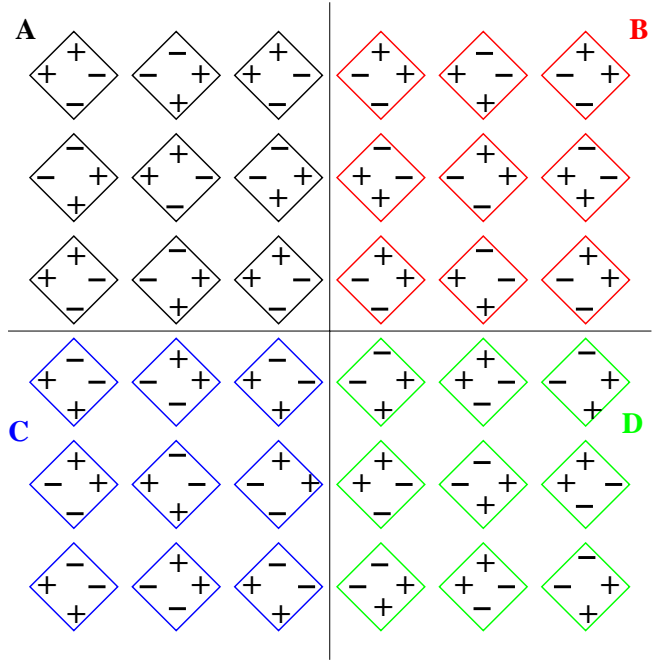}
\caption{\label{D110} (Color online)
Domain walls in CMO for $T_3>T>{\tilde T}_2$.
Here one has low energy 90$^{\rm o}$ walls between phases A and B, 
B and D, and C and D.  Antiphase regions A and D are separated by
two such walls.  If the distance between regions A and D is not
too large, it may be possible to establish the existence of phase
domains using the technique of Ref. \onlinecite{BODE}.}
\end{center}
\end{figure}

\noindent
Upon cooling through $T_3$ it seems likely
that one would have different nucleation sites from which ordering
would develop.  As a result, one would have randomly chosen signs of
the OP $Q_3({\bf q})$ in different regions of the sample.
The question then arises, would these domains coarsen and
the sample then become a single domain sample?  To answer this
question we need to study the energetics of domain walls.
This involves understanding structural defects, as were studied
in Refs. \onlinecite{DAVIES} and \onlinecite{ARIAS}.  
For this discussion, I first consider the simpler case of an $n=1$
RP system for which the tilt is around a [100] direction.
In Fig. \ref{DOM1} a domain wall is shown separating two phases to
the left and right of the wall which differ only in their phase.
One sees that if the domain wall is a perfect plane,
then there will be no intraoctahedral energy involved in the
wall.  Indeed, the only energy of the domain wall will be that
from the potential that keeps the plane in its location.  However,
as explained in the caption to Fig. \ref{DOM1}, kink formation requires
a large energy.  Notice the difference between domain walls in
this system and those in an Ising antiferromagnet.  In an Ising
antiferromagnet the domain wall energy is proportional to the length
of the wall.  Having a kink simply increases the energy by one
unit of exchange energy.  Here a kink either has the nonlocal
string-like energy of a kink-antikink or its has the large energy
needed to deform the octahedra.  Thus, it seems that domain wall
motion may be inhibited and therefore it is possible that phase 
domains, once created will remain in the sample.

From these examples one concludes that a domain wall tends to form in 
planes of ``minimum contact," {\it i. e.} in planes which intersect
the least number of shared oxygen ions.  This is why $[100]$ walls are
preferred over $[110]$ walls for systems, such as CMO, in which tilting
occurs about a $[110]$ direction.[\onlinecite{DAVIES}] Of course, $[001]$
walls (stacking faults) probably have the least energy.

\section{CONCLUSIONS}

In this paper we have explored the rich structure of structural,
magnetic, and dielectric ordering in the Ruddlesden-Popper compound
Ca$_3$Mn$_2$O$_7$, using Landau theory to analyze symmetry properties.  Our
approach is similar to that used by Perez-Mato et al. [\onlinecite{PEREZ}]
to study the Aurivillius compounds.  Most of the symmetry relations
we find are explicitly corroborated by the first principles calculations
of Benedek and Fennie.  An important aspect of our work is
to motivate a large number of experiments which can elucidate the
relations between the various order parameters. Specifically,
we summarize the conclusions from our work as follows

$\bullet \ 1$ 
The most important aspect of our work is that we introduce order
parameters (OP's) for all the irreducible representations (irreps)
for all the wave vectors of the star which is active in the
ordering transitions.  The OP's describe distortions from the
parent high symmetry tetragonal lattice which exists at high
temperature.  This enables us to discuss the induced
(nonprimary) OP's such as the spontaneous polarization,
the weak ferromagnetism, and the elastic strains.

$\bullet  2$
In conformity with established results[\onlinecite{UTAH}] (but
rejecting multicritical points) we treat the group-subgroup
structure obtained by the first principles calculations of
Benedek and Fennie.[\onlinecite{NATURE}]  We give an OP description of
the sequence of structural transitions which is predicted for
Ca$_3$Mn$_2$O$_7$ and Ca$_3$Ti$_2$O$_7$,
namely I4/mmm $\rightarrow$ Cmcm $\rightarrow$ Cmc2$_1$ the same as that
found for a similar perovskite by Perez-Mato {\it et al.}.[\onlinecite{PEREZ}]

$\bullet \ 3$
The ordering involves two families of domains, one family for each
of the two X wave vectors.  At the lower structural transition, a
ferroelectric polarization appears parallel to the wave vector.  Below that
there is an independent magnetic ordering transition to an
antiferromagnetic state in which the stacking of the magnetic bilayers
depends on the wave vector which was selected when the tetragonal symmetry
was broken.  A weak ferromagnetic moment develops perpendicular to both
the staggered magnetization and the polarization.

$\bullet 4$ 
We show that by application of an applied
magnetic field it might be possible to reorient the ferromagnetic
moment through successive 90$^{\rm o}$ rotations, which would then
induce similar rotations of the spontaneous polarization.  Likewise,
application of an external electric field could reorient the
spontaneous polarization which in turn would reorient the wave
vector  and thereby reorient the weak ferromagnetic moment.

$\bullet 5$ 
Here we analyzed behavior near the phase transitions using mean field 
theory.  However, non mean-field critical exponents can be accessed
experimentally, as has been done for Ni$_3$V$_2$O$_8$.[\onlinecite{GLNVO1}]

$\bullet \ 6$
We have given a detailed enumeration (see Fig. \ref{DOMFIG}) of the
domains arising from different realization of the OP's.  We
distinguish between domains whose bulk structure is macroscopically
identifiable and those (similar to antiferromagnetic domains) 
that arise from a difference in phase that is not macroscopically 
accessible.  We propose the association of domain walls with planes
of ``minimum contact" between octahedra.  The question of the formation
and dynamics of domains (especially those of phase domains) is broached.

\noindent{\bf Acknowledgments}
I would like to thank C. J. Fennie for introducing me to this subject and
for useful advice.  I also thank M. V. Lobanov, H. T. Stokes, B. Campbell,
J. M. Perez-Mato, and J. Kikkawa for helpful discussions.

\begin{appendix}

\section{Crystal Structure for $Q_2^+({\bf q}_1)$ and $Q_3^-({\bf q}_1)$}
\label{APPA}

Here we verify that the crystal structure when the OP's for ireps $X_2^+$
and $X_3^-$ at wave vector ${\bf q}_1$ are simultaneously nonzero is Cmc2$_1$.  
(This result also applies when the OP's are both at wave vector ${\bf q}_2$.)
From Table \ref{IRREP} we have the characters listed in Table \ref{TAPP}.

Now which operators transform like unity under both irreps?
We see that we may choose
\begin{eqnarray}
T_1T_2 \ , \hspace{0.15 in} 
T_1T_2^{-1} \ , \hspace{0.15 in} 
T_1^{-1}T_3 \ , \hspace{0.15 in} 
m_z \ , \hspace{0.15 in} 
m_d T_1 \ .
\nonumber \end{eqnarray}
These indicate that we new primitive lattice vectors are
\begin{eqnarray}
{\bf a}_1 &=& (1,1,0) \ , \hspace{0.15 in}
{\bf a}_2 = (1,\overline 1,0) \ , \hspace{0.15 in}
{\bf a}_3 = (-1/2,1/2,1/2) \ .
\nonumber \end{eqnarray}
Also
\begin{eqnarray}
m_d &=& (x,y,z,) \rightarrow (x,y, \overline z) \ , \nonumber \\
m_d T_1 &=& (x,y,z) \rightarrow (y,x+1,z)
\nonumber \end{eqnarray}

\begin{table} [h!]
\caption{\label{TAPP} Characters of generators of the little group for
irreps $X_2^+$ and $X_3^-$ at wave vector ${\bf q}_1$.}

\vspace{0.2 in}
\begin{tabular} {||c | c c c c c c||}
\hline \hline
${\cal O}=$ & ${\cal I}$ & $m_d$ & $m_z$ & $T_1$ & $T_2$ & $T_3$ \\
\hline
$X_2^+$ & 1 & $-1$ & 1 & $-1$ & $-1$& $-1$ \\ 
$X_3^-$ & $-1$ & $-1$ & 1 & $-1$ & $-1$& $-1$ \\ 
\hline \hline
\end{tabular}
\end{table}

\noindent
To make contact with Ref. \onlinecite{ITC} we transform to orthorhombic 
coordinates:
\begin{eqnarray}
x' &=& z \ , \hspace{0.15 in}
y' = \frac{x-y}{2} + \frac{1}{4}  \ , \hspace{0.15 in}
z' = \frac{x+y}{2} \ .
\nonumber \end{eqnarray}
In this coordinate system 
\begin{eqnarray}
{\bf a}_1' &=& (0,0,1) \ , \hspace{0.15 in}
{\bf a}_2' = (0,1,0) \ , \hspace{0.15 in}
{\bf a}_3' = (1/2,-1/2,0) \ , 
\nonumber \end{eqnarray}
and the mirror operations are
\begin{eqnarray}
m_x' &=& (x',y',z') \rightarrow (\overline x',y',z') \nonumber \\
\ [m_dT_1]' &=& (x',y',z') \rightarrow (x',\overline y', z'+1/2)
\nonumber \end{eqnarray}
which coincides with the specification of space group Cmc2$_1$ in
Ref. \onlinecite{ITC}. One might object that we have not taken
into account the fact that Eq. (\ref{CUBIC}) indicates the presence
of irrep $\Gamma_5^-$.  What that means is that this irrep is always
allowed in Cmc2$_1$.

\section{CRITICAL BEHAVIOR OF THE DIELECTRIC CONSTANT}
\label{APPC}

The following discussion parallels that given[\onlinecite{R2}] for
the dielectric anomaly in Ni$_3$V$_2$O$_8$.  We take the free energy
(in the ${\bf q}_1$ channel) for $T$ near the lower transition 
[where $Q_3({\bf q}_1)$ is already nonzero] to be
\begin{eqnarray}
{\cal F} &=& \frac{a'}{2} (T-T_2) Q_2({\bf q}_1)^2 
+ \frac{a''}{2} (T-T_5) [Q_{5,1}+ Q_{5,2}]^2 \nonumber \\
&& \ + w' Q_2({\bf q}_1) [Q_{5,1}+ Q_{5,2}]
- \sqrt 2 \lambda [P_xQ_{5,1}+P_yQ_{5,2}] 
\nonumber \\ && \ + \frac{1}{2} \chi_E^{-1} {\bf P}^2
- \lambda' \langle Q_3 ({\bf q}_1) \rangle
Q_2({\bf q}_1) [P_x+P_y]
\nonumber \\ && \  - [P_xE_x + P_y E_y] \ .
\label{B1} \end{eqnarray}
The coupling terms proportional to $\lambda$ and $\lambda'$ suggest
that the dielectric susceptibility will be singular at the
lower transition where $Q_2$ and $Q_5$ appear.  
In this appendix we show this explicitly.  We write
\begin{eqnarray}
P_x Q_{5,1} + P_y Q_{5,2} &=& \frac{1}{2} [P_x+P_y][Q_{5,1}+Q_{5,2}]
\nonumber \\ && \ + \frac{1}{2} [P_x-P_y][Q_{5,1}-Q_{5,2}]
\nonumber \end{eqnarray}
and
\begin{eqnarray}
P_xE_x + P_y E_y &=& \frac{1}{2} [P_x+P_y][E_x+E_y]
\nonumber \\ && \ + \frac{1}{2} [P_x-P_y][E_x-E_y] \ .
\nonumber \end{eqnarray}
Note that in the ${\bf q}_1$ channel $Q_{5,1}-Q_{5,2}=0$ and can be dropped.
Also, since $E_x -E_y$ and $P_x-P_y$ do not couple to a critical variable
we drop them too.  Now set 
\begin{eqnarray}
P & \equiv & [P_x+P_y]/\sqrt 2 \ , \hspace{0.25 in}
E  \equiv  [E_x+ E_y]/\sqrt 2 \ , \nonumber \\ && \
Q_5 = Q_{5,1} + Q_{5,2} \ .
\nonumber \end{eqnarray}
Thus the above free energy is
\begin{eqnarray}
{\cal F} &=& \frac{a'}{2} (T-T_2) Q_2^2 
+ \frac{a''}{2} (T-T_5) Q_5^2 + w' Q_2 Q_5 \nonumber \\ &&
+ \frac{1}{2} \chi_E^{-1} P^2
-  \lambda P Q_5 - \lambda'' Q_2 P - PE \ ,
\nonumber \end{eqnarray}
where $\lambda'' = \lambda' \langle Q_3({\bf q}_1) \rangle \sqrt 2$.
Now minimize with respect to $Q_2$ and $Q_5$ to get
\begin{eqnarray}
\frac{\partial {\cal F}}{\partial Q_2} &=& a' (T-T_2) Q_2 + w' Q_5
- \lambda'' P = 0 \ , \nonumber \\
\frac{\partial {\cal F}} {\partial Q_5} &=& w'Q_2 + a'' (T-T_5) Q_5
- \lambda P = 0 \ ,
\nonumber \end{eqnarray}
so that
\begin{eqnarray}
Q_2 &=& [\lambda'' a'' (T-T_5) -w' \lambda]P/D \nonumber \\
Q_5 &=& [-\lambda'' w' +  a' (T-T_2) \lambda]P/D \ ,
\nonumber \end{eqnarray}
where
\begin{eqnarray}
D &=& a' (T-T_2) a'' (T-T_5) - {w'}^2 \ .
\nonumber \end{eqnarray}
Then the equation for $P$ is $\partial {\cal F}/ \partial P =0$, or
\begin{eqnarray}
- \lambda'' Q_2 - \lambda Q_5 + \chi_E^{-1} P = E \ ,
\nonumber \end{eqnarray}
so that the dielectric susceptibility is
\begin{eqnarray}
{\tilde \chi}_{E} & \equiv & \frac{P}{E} =\nonumber \\ &=&
D \left[ a' a'' \chi_E^{-1} (T-T_2)(T-T_5) - {w'}^2 \chi_E^{-1} \right.
\nonumber \\ && \  \left. - \lambda^2 a'(T-T_2)
- {\lambda''}^2 a'' (T-T_5) + 2 w' \lambda \lambda'' \right]^{-1} \ .
\nonumber \end{eqnarray}
Note that ${\tilde \chi}_E$ has poles at $T={\tilde T}_2$  (which is close to
$T_2$) and at ${\tilde T}_5$ (which is close to $T_5$). When the effects of
$w'$, $\lambda$, and $\lambda''$ can be treated perturbatively with respect
to $T_2$ and $\chi_E^{-1}$, we find that
\begin{eqnarray}
{\tilde T}_2 &=& T_2 + \frac{{w'}^2}{a'a''T_2}
+ \frac{{\lambda''}^2\chi_E}{a'} \equiv T_2 + \delta T_2
\nonumber \end{eqnarray}
a result which is reasonable considering the couplings proportional to
$w'$ and $\lambda'$ in Eq. (B1).  For $T$ near ${\tilde T}_2$ we
can therefore write
\begin{eqnarray}
{\tilde \chi}_E &=& \frac{D \chi_E}{a'a''(T-{\tilde T}_2)( T - {\tilde T}_5)}
=  \frac{A \chi_E}{T- {\tilde T}_2} \ ,
\nonumber \end{eqnarray}
where, in terms of $D(T)$, we have 
\begin{eqnarray}
A \approx \frac{D(T={\tilde T}_2)}{a'a''T_2} = \frac{a'a''T_2 (\delta T_2)-
{w'}^2}{a'a''T_2} = \frac{{\lambda''}^2 \chi_E}{a'} \ ,
\nonumber \end{eqnarray}
which is a small amplitude attributable to the existence of the mixing term
proportional to $\lambda'$ in Eq. (\ref{B1}).  But this result does confirm the
expected divergence in the dielectric constant at the lower transition
where the polarization first appears.
\end{appendix}


\begin{thebibliography} {99}
\bibitem{FOOT}
Probably BF find that when the spin-orbit interaction is included
the spins point along the tetragonal [001] direction as they state.
However, there is some confusion, possibly notational, when they
state (in the previous sentence) that the polarization is along [010],
which would be true if this were in orthorhombic coordinates
and ${\bf q}$ were appropriately chosen.
\bibitem{FN1}
These irreps are irreps of the full space group for the star of the wave
vector.  For $\Gamma_5^-$ this irrep is the same as that for the group
of the wave vector.  For $X_2^+$ and $X_3^-$ the matrices for the operators
of the group of the wave vector are block diagonal, i. e. they form a
reducible representation.  However, matrices for operators which are in
the full space group but which are {\it not} in the group of the wave vector
are not diagonal.  Therefore this irrep contains information on how the
basis functions transform under the operator ${\cal R}_4$ which is not a
member of the group of the wave vector. The first row and column of
the matrices refer to ${\bf q}_1$ and the second ones to ${\bf q}_2$.
\bibitem{FNA}
We put a superscript $+$ or $-$ on the OP to indicate its parity under
inversion when that information needs emphasis.
\bibitem{EIGHT}
In BF the number of nonmagnetic domains is quoted as being four.  Apparently
they did not count the two choices of wave vector.
\bibitem{FN3}
The first term in Eq. (23) may not accurately follow a power law
for $T$ near ${\tilde T}_2$, but the point we make is that there
is an anomaly due to the second term.
\end{thebibliography}
\end{document}